\documentclass[
aps,prd,
12pt,%10pt
nopreprintnumbers,
showpacs,
eqsecnum,
nofootinbib
]{revtex4-1}

\usepackage{graphicx}
\usepackage{amssymb}

\begin{document}

\title{Vacuum expectation values in non-trivial background space
from three types of UV improved Green's functions}
\author{Nahomi Kan}\email[]{kan@gifu-nct.ac.jp}
\affiliation{National Institute of Technology, Gifu College,
Motosu-shi, Gifu 501-0495, Japan
}
\author{Masashi Kuniyasu}\email[]{mkuni13@yamaguchi-u.ac.jp}
\author{Kiyoshi Shiraishi}\email[]{shiraish@yamaguchi-u.ac.jp}
%\author{Kohjiroh Takimoto}\email[]{i016vb@yamaguchi-u.ac.jp}
\author{Zhenyuan Wu}\email[]{b501wb@yamaguchi-u.ac.jp}
\affiliation{
Graduate School of Sciences and Technology for Innovation, Yamaguchi
University, Yamaguchi-shi, Yamaguchi 753--8512, Japan}
\date{\today}
%\date{}

\begin{abstract}
We evaluate the quantum expectation values in non-simply connected spaces
by using
UV improved Green's functions proposed by Padmanabhan, Abel, and Siegel.
It is found that the results from these three types of Green's functions behave
similarly under changes of scales, but have minute differences.
Prospects in further applications are briefly discussed.
\end{abstract}

%\preprint{}

\pacs{%
%02.10.Ox, %%%Combinatorics; graph theory
%02.20.Sv, %Lie algebra of Lie groups
%02.30.Hq, %Ordinary differential equations
%02.30.Ik, %Integrable systems
%02.30.Jr, %Partial differential equations
%02.40.Gh, %Noncommutative geometry
%03.65.-w, %Quantum mechanics
%03.65.Db, %Functional analytical methods
%03.65.Sq, %Semiclassical theories in quantum mechanics
03.70.+k, %Theory of quantized fields
%04.20.-q, %%%Classical general relativity
%04.20.Fy, %%Canonical formalism, Lagrangians, and variational principles
%04.20.Jb, %%Exact solutions
%04.25.-g, %Approximation
%04.25.Nx, %%%Post-Newtonian approximation; perturbation theory; related
%approximations
%04.40.-b, %Self-Gravitating systems
%04.40.Nr, %%Einstein-Maxwell spacetime
%04.50.-h, %%%%%Higher-dimensional gravity and other theories of gravity 
%04.50.Cd, %Kaluza-Klein theories 
%04.50.Gh, %Higher-dimensional black holes, black strings, 
%and related objects 
%04.50.Kd, %%%Modified theories of gravity 
04.60.-m, %%Quantum gravity
%04.60.Kz, %%Lower dimensional models; minisuperspace models
%04.60.Rt, %Topologically massive gravity
%04.62.+v, %Quantum fields in curved spacetime
%04.65.+e, %Supergravity
%04.70.Bw, %%%Classical black holes
%05.30.Jp, %Boson systems
%11.10.-z, %%%Field theory
%11.10.Lm, %%%Nonlinear or nonlocal theories and models 
%11.10.Nx, %%%Noncommutative field theory 
%11.10.Kk, %%%Field theories in dimensions other than four
%11.25.-w, %Strings and branes
11.25.Mj, %%Compactification and four-dimensional models
11.27.+d% %%Extended classical solutions; cosmic strings, 
%domain walls, texture 
%11.30.-j, %Symmetry and conservation laws
%11.30.Pb, %Supersymmetry
%12.60.-i, %Models beyond the standard model
%45.20.Jj, %Lagrangian and Hamiltonian mechanics
%95.35.+d, %Dark matter
%95.36.+x, %Dark energy
%98.80.-k, %%%Cosmology 
%98.80.Cq, %%%%%Particle-theory and field-theory models of the early
%Universe  
%98.80.Dr, %Relativistic cosmology 
%98.80.Qc, %Quantum cosmology
%98.80.Jk% %%Mathematical and relativistic aspects of cosmology
.}

\maketitle

%%%%%%%%%%%%%%%%%%%%%%%%%%%%%%%%%%%%%%%%%%%%%%%%%%%%%%%%%%%%%%%%%%%%%%%%%%%
%%%%%%%%%%%%%%%%%%%%%%%%%%%%%%%%%%%%%%%%%%%%%%%%%%%%%%%%%%%%%%%%%%%%%%%%%%%
%%%%%%%%%%%%%%%%%%%%%%%%%%%%%%%%%%%%%%%%%%%%%%%%%%%%%%%%%%%%%%%%%%%%%%%%%%%
\section{Introduction}
\label{introduction}
%%%%%%%%%%%%%%%%%%%%%%%%%%%%%%%%%%%%%%%%%%%%%%%%%%%%%%%%%%%%%%%%%%%%%%%%%%%
%%%%%%%%%%%%%%%%%%%%%%%%%%%%%%%%%%%%%%%%%%%%%%%%%%%%%%%%%%%%%%%%%%%%%%%%%%%
%%%%%%%%%%%%%%%%%%%%%%%%%%%%%%%%%%%%%%%%%%%%%%%%%%%%%%%%%%%%%%%%%%%%%%%%%%%

The appearance of divergences in the ultra-violet (UV) region is one of the
fundamental problems in quantum field theory.
The problem of UV completion is expected to be connected to quantum gravity,
because the UV divergence is naively considered to be related to fine 
structures of spacetime.

Almost two decades ago, Padmanabhan \cite{Pad1,Pad2,Pad3,Pad4,Pad5,Pad6} and his
collaborators \cite{SSP,KSSP} considered a UV completed propagator involving
the ``Planck length'' as a cutoff small-length scale. Padmanabhan's propagator was
induced by the duality in the integrand kernel with a proper-time parameter.
More recently, Abel and his collaborators \cite{AD,ABM} proposed their UV
improved propagator, which is also inspired by the integral over modular parameter
in string theory. Modification of integration measure or range was also offered by
Siegel \cite{Siegel} at the beginning of this century, motivated by an infinite
derivative theory.  

The authors who mentioned UV improved propagators studied and calculated various
physical quantities mainly in homogeneous Minkowski
spacetime.\footnote{Only Padmanabhan and his collaborators also considered thermal
effects in the Rindler space as well as quantum effects including Casimir effect
in Ref.~\cite{SSP}, and quantum effects in the constant-curvature space time in
Ref.~\cite{KSSP}. } Because the UV behavior is believed to be concerned with
quantum gravity, it will be important to study further on physical results of
the UV completion in non-trivial spacetimes.

The non-trivial space that can be handled most easily is a non-simply
connected space. In the present paper, Euclidean Kaluza--Klein space and space
with a conical singularity are adopted as background spaces.%
\footnote{Thus, in the present paper, we do not deal with the problem on the UV
behavior of the propagator in the spacetime with the Lorentz signature
(see, for example, Ref.~\cite{BLM}).}  It is already known that the vacuum energy
density and the stress tensor of a scalar field can be obtained from the congruent
limit of the propagator. We will examine how the ``Planck length'', which is used
as a common synonym for the fundamental cutoff length, works for the quantum
quantities and compare the consequences arising from the three modified {\it
Green's functions} (we use this term because we use the Euclidean metric in this
paper).

This paper is organized as follows.
The next section contains a review of the Green's function for a massless scalar
field and an introduction of three UV improved Green's functions,
say, Padmanabhan's type, Abel's type, and Siegel's type.
They can be expressed by the integral form of the heat kernel of the Laplacian.
In Sec.~\ref{sec3}, we consider a partially compactified space \`a la Kaluza-Klein.
We calculate the vacuum energy density using three UV improved
Green's functions in the non-simply connected space $\mathbf{R}^{D-1}\otimes S^1$.
The dependence of the energy densities on the ratio of the circumference of
$S^1$ and the Planck length is shown.  In Sec.~\ref{sec4} the expectation values
for the stress tensor are computed for the three types of Green's functions in a
conical space. Finally, in Sec.~\ref{conclusion} we
conclude with a discussion of future directions.
At the end of this paper, we attach Appendices \ref{AA} and \ref{AB}, where
mathematical definitions and formulas are gathered, for convenience.

%%%%%%%%%%%%%%%%%%%%%%%%%%%%%%%%%%%%%%%%%%%%%%%%%%%%%%%%%%%%%%%%%%%%%%%%%%%
%%%%%%%%%%%%%%%%%%%%%%%%%%%%%%%%%%%%%%%%%%%%%%%%%%%%%%%%%%%%%%%%%%%%%%%%%%%
%%%%%%%%%%%%%%%%%%%%%%%%%%%%%%%%%%%%%%%%%%%%%%%%%%%%%%%%%%%%%%%%%%%%%%%%%%%
\section{Three UV improved Green's functions and the heat kernels in homogeneous
and isotropic Euclidean space $\mathbf{R}^D$}
\label{sec2}
%%%%%%%%%%%%%%%%%%%%%%%%%%%%%%%%%%%%%%%%%%%%%%%%%%%%%%%%%%%%%%%%%%%%%%%%%%%
%%%%%%%%%%%%%%%%%%%%%%%%%%%%%%%%%%%%%%%%%%%%%%%%%%%%%%%%%%%%%%%%%%%%%%%%%%%
%%%%%%%%%%%%%%%%%%%%%%%%%%%%%%%%%%%%%%%%%%%%%%%%%%%%%%%%%%%%%%%%%%%%%%%%%%%

Let us first recall some basic facts with standard Green's functions and heat
kernels of a flat metric \cite{Vassilevich,Camporesi}.
Suppose that the standard Green's function
 $G_D(x,x')$ of a massless scalar
field in a $D$-dimensional Euclidean space $\mathbf{R}^D$ satisfies
\begin{equation}
-\Delta_x{G}_D(x,x')=\frac{1}{\sqrt{g}}\delta^D(x,x')\,,
\end{equation}
where $\Delta=\frac{1}{\sqrt{g}}\partial_\mu\sqrt{g}g^{\mu\nu}\partial_\nu$ is the
Laplacian operator acting on scalar fields and
$\frac{1}{\sqrt{g}}\delta^D(x,x')$ is the $D$-dimensional covariant delta function.

The main tool that we use here is the heat kernel of the Laplacian.
The heat kernel $K_D(x,x';s)$ is introduced as a representation of the
Green's function written by using it:
\begin{equation}
{G}_D(x,x')=\int_0^\infty ds\, K_D(x,x';s)\,.
\end{equation}
Here, the heat kernel obeys the equation
\begin{equation}
\left[\frac{\partial}{\partial s}-\Delta_x\right]K_D(x,x';s)=0\,,
\end{equation}
with the boundary condition $\lim_{s\rightarrow
0}K_D(x,x';0)=\frac{1}{\sqrt{g}}\delta^D(x,x')$. Subsequently, 
the heat kernel in
$\mathbf{R}^D$ is found to be
\begin{equation}
K_D(x,x';s)=\int\frac{d^Dp}{(2\pi)^D} \exp\left[-s p^2\right]e^{ip\cdot
(x-x')} =\frac{1}{(4\pi s)^{D/2}}\exp\left({-\frac{w^2}{4s}}\right)\,,
\end{equation}
and the massless Green's function reads
\begin{equation}
G_D(x,x')=\frac{\Gamma\left({\textstyle
\frac{D-2}{2}}\right)}{4\pi^{D/2}w^{D-2}}\equiv G_D(w)\,,
\end{equation}
where $w\equiv \sqrt{\scriptstyle(x-x')\cdot(x-x')}$ and $\Gamma(z)$ is the gamma
function.

Next, we consider UV improved heat kernels and Green's functions.
Padmanabhan advocated a UV modified heat kernel
\cite{Pad1,Pad2,Pad3,Pad4,Pad5,Pad6}
\begin{equation}
K^{(P)}_D(x,x';s)=K_D(x,x';s)\exp\left(-\frac{l_p^2}{4s}\right)\,,
\end{equation}
where $l_p$ is interpreted as the Planck length.
Obviously, Padmanabhan's heat kernel leads to the Green's function $G_D^{(P)}(w)$
in
$\mathbf{R}^D$, which can be written as
\begin{equation}
G_D^{(P)}(w)=G_D\left({\textstyle \sqrt{w^2+l_p^2}}\right)\,.
\end{equation}
This shows the simplest introduction of the fundamental length $l_p$
in the Green's function of Padmanabhan's type and it give rise to the UV
completion.

Recently, Abel et al.~considered another UV modified heat kernel \cite{AD,ABM},
which can be regarded as\footnote{The fundamental length $l_p$ can be freely chosen
for each type of the kernel. For simplicity, we exploit a common one to describe
all three cases.}
\begin{equation}
K^{(A)}_D(x,x';s)=K_D(x,x';{\textstyle s+\frac{l_p^4}{4s}})\,.
\end{equation}
In this case, the Green's function can be expressed as 
\begin{equation}
G^{(A)}_D(x,x')=\int_{l_p^2}^\infty dT \frac{T}{\sqrt{T^2-l_p^4}}K_D(x,x';T)\,,
\end{equation}
after the change of integration variable.
Performing the integral
yields a
complicate form for the massless Green's function of Abel's type in $\mathbf{R}^D$:
\begin{eqnarray}
G^{(A)}_D(x,x')&=&\frac{1}{2^D\pi^{(D-1)/2}l_p^{D-2}}\left[
\frac{\Gamma({\textstyle\frac{D-2}{4})}}{2\Gamma({\textstyle\frac{D}{4}})}
\,{}_1F_2({\textstyle\frac{D-2}{4};\frac{1}{2},\frac{D}{4};\frac{w^4}{64l_p^4}})
-\right.
\nonumber \\
& &\qquad\qquad\qquad\qquad\left.
\frac{w^2}{8l_p^2}
\frac{\Gamma({\textstyle\frac{D}{4}})}{\Gamma({\textstyle
\frac{D+2}{4}})}\,{}_1F_2({\textstyle\frac{D}{4};\frac{3}{2},
\frac{D+2}{4};\frac{w^4}{64l_p^4}})\right]\,,
\label{cif}
\end{eqnarray}
where ${}_pF_q$ is the hypergeometric function.
For $D=4$, the expression (\ref{cif}) reads
\begin{equation}
G^{(A)}_4(x,x')=\frac{1}{32\pi l_p^2}\left[I_{0}({\textstyle
\frac{w^2}{4l_p^2}})-
\mathbf{L}_{0}({\textstyle\frac{w^2}{4l_p^2}})\right]\,,
\end{equation}
where $I_\nu(z)$ is the modified Bessel function of the first kind and
$\mathbf{L}_\nu(z)$ is the modified Struve function (see Appendix \ref{AA}).

Siegel suggested a simple UV completion \cite{Siegel}, which is achieved by
\begin{equation}
K^{(S)}_D(x,x';s)=K_D(x,x';s+l_p^2)\,.
\end{equation}
Then, the Siegel-type Green's function for a massless scalar field in
$\mathbf{R}^D$ reads
\begin{equation}
G^{(S)}_D(x,x')=\frac{1}{4\pi^{D/2}w^{D-2}}\gamma
({\textstyle \frac{D-2}{2}, \frac{w^2}{4l_p^2}})
=\frac{1}{4\pi^{D/2} w^{D-2}}\left[\Gamma
({\textstyle \frac{D-2}{2}})
-
\Gamma
({\textstyle \frac{D-2}{2}, \frac{w^2}{4l_p^2}})
\right]\,,
\end{equation}
where $\gamma(z,\alpha)$ and $\Gamma(z,\alpha)$ are the incomplete gamma functions
which are defined by $\gamma(z,\alpha)\equiv\int_0^\alpha t^{z-1}e^{-t} dt$ and
$\Gamma(z,\alpha)\equiv\int_\alpha^\infty t^{z-1}e^{-t} dt$, respectively.

Obviously, taking the limit of $l_p\rightarrow 0$, we find
\begin{equation}
\lim_{l_p\rightarrow 0}G^{(P)}_D(x,x')=
\lim_{l_p\rightarrow 0}G^{(A)}_D(x,x')=
\lim_{l_p\rightarrow 0}G^{(S)}_D(x,x')=
G_D(x,x')\,.
\end{equation}

In the opposite limit, $x'\rightarrow x$, we get finite results immediately as
\begin{eqnarray}
\lim_{x'\rightarrow x}G^{(P)}_D(x,x')&=&\frac{\Gamma\left(
{\textstyle\frac{D-2}{2}}\right)}{4\pi^{D/2}l_p^{D-2}}\,,\nonumber \\
\lim_{x'\rightarrow
x}G^{(A)}_D(x,x')&=&
\frac{\Gamma({\textstyle\frac{D-2}{4}})}{2^{D+1}\pi^{(D-1)/2}
\Gamma({\textstyle\frac{D}{4}})l_p^{D-2}}\,,\nonumber
\\
\lim_{x'\rightarrow x}G^{(S)}_D(x,x')&=&\frac{1}{2^{D-1}(D-2)\pi^{D/2}l_p^{D-2}}\,,
\label{acs}
\end{eqnarray}
and find that they are $O(l_p^{2-D})$ as expected.

In the following sections, we generalize three UV improved Green's functions
to treat quantum quantities in non-simply connected space.

%%%%%%%%%%%%%%%%%%%%%%%%%%%%%%%%%%%%%%%%%%%%%%%%%%%%%%%%%%%%%%%%%%%%%%%%%%%
%%%%%%%%%%%%%%%%%%%%%%%%%%%%%%%%%%%%%%%%%%%%%%%%%%%%%%%%%%%%%%%%%%%%%%%%%%%
%%%%%%%%%%%%%%%%%%%%%%%%%%%%%%%%%%%%%%%%%%%%%%%%%%%%%%%%%%%%%%%%%%%%%%%%%%%
\section{UV improved Green's functions and vacuum energy density in a partially
compactified space
$\mathbf{R}^{D-1}\otimes S^1$}
\label{sec3}
%%%%%%%%%%%%%%%%%%%%%%%%%%%%%%%%%%%%%%%%%%%%%%%%%%%%%%%%%%%%%%%%%%%%%%%%%%%
%%%%%%%%%%%%%%%%%%%%%%%%%%%%%%%%%%%%%%%%%%%%%%%%%%%%%%%%%%%%%%%%%%%%%%%%%%%
%%%%%%%%%%%%%%%%%%%%%%%%%%%%%%%%%%%%%%%%%%%%%%%%%%%%%%%%%%%%%%%%%%%%%%%%%%%
We consider a $D$-dimensional space whose metric is given by
\begin{equation}
ds^2=\sum_{i=1}^{D-1}dz_i^2+dy^2\,,
\end{equation}
where the coordinate $y$ is the coordinate on a circle ($S^1$) and we
assume that $0\le y<L$, where $L$ is the circumference of the circle.
The periodic boundary condition in the direction of
$y$ is supposed to be applied on the massless scalar, i.e., $\phi(y+L)=\phi(y)$.
In this space, $\mathbf{R}^{D-1}\otimes S^1$, the heat kernel for the standard
massless scalar field takes the form
\begin{eqnarray}
K_D(z,z',y,y';s)&=&\sum_{n=-\infty}^{\infty}\frac{1}{L}\int\frac{d^{D-1}p}{(2\pi)^{D-1}}
\exp\left[-s\left({\textstyle p^2+\frac{4\pi^2n^2}{L^2}}\right)\right]e^{ip\cdot
(z-z')+i2\pi n(y-y')/L}
\nonumber \\
&=&\frac{1}{(4\pi
s)^{D/2}}\sum_{n=-\infty}^{\infty}\exp\left[-\frac{\zeta^2+(\eta-nL)^2}{4s}\right]
\equiv K_D(\zeta,\eta;s)\,,
\end{eqnarray}
where $\zeta=\sqrt{\scriptstyle(z-z')\cdot(z-z')}$ and $\eta=y-y'$.
The second expression with the periodic summation can be derived with the aid of
the formula of the theta function (see Appendix \ref{AA}) as well as from the
method of images. We should also notice that the limit
$L\rightarrow\infty$ yields the heat kernel in $\mathbf{R}^D$.

In the subsections below, we study the Green's functions of Padmanabhan-type,
Abel-type, and Siegel-type in $\mathbf{R}^{D-1}\otimes S^1$, and the
vacuum energies extracted from them.

%%%%%%%%%%%%%%%%%%%%%%%%%%%%%%%%%%%%%%%%%%%%%%%%%%%%%%%%%%%%%%%%%%%%%%%%%%%
%%%%%%%%%%%%%%%%%%%%%%%%%%%%%%%%%%%%%%%%%%%%%%%%%%%%%%%%%%%%%%%%%%%%%%%%%%%
\subsection{Vacuum energy from Padmanabhan-type Green's function}
%%%%%%%%%%%%%%%%%%%%%%%%%%%%%%%%%%%%%%%%%%%%%%%%%%%%%%%%%%%%%%%%%%%%%%%%%%%
%%%%%%%%%%%%%%%%%%%%%%%%%%%%%%%%%%%%%%%%%%%%%%%%%%%%%%%%%%%%%%%%%%%%%%%%%%%

Padmanabhan-type Green's function in $\mathbf{R}^{D-1}\otimes S^1$ is
found to be
\begin{equation}
K_D^{(P)}(\zeta,\eta;s)=K_D({\textstyle
\sqrt{\zeta^2+l_p^2}},\eta;s)\,.
\end{equation}
We can calculate the effective action $W$ from the trace of heat kernel.
In the present case, the effective action in terms of Padmanabhan's heat
kernel becomes
\begin{equation}
W_D^{(P)}=\frac{1}{2}L\int d^{D-1}z \int_0^\infty \frac{ds}{s} K_D^{(P)}(0,0;s)
=\frac{1}{2}L\int d^{D-1}z \int_0^\infty \frac{ds}{s} K_D(l_p,0;s)\,.
\end{equation}
It is noteworthy that, because of the UV modification, a cutoff in $s$-integration
or an additional power of $s$ in the integrand is unnecessary.

Consequently, the one-loop vacuum energy density $U_D^{(P)}=-W_D^{(P)}/V_{D-1}$,
where $V_{D-1}$ is the
$(D-1)$-dimensional volume, reads
\begin{eqnarray}
U_D^{(P)}(L)&=&-\frac{1}{2\pi^{D/2}L^{D-1}}\int_0^\infty dt\,
t^{D/2-1}\sum_{n=-\infty}^{\infty}
\exp\left[-\left({\textstyle n^2+\frac{l_p^2}{L^2}}\right)t\right]\nonumber \\
&=&-\frac{1}{2\pi^{D/2}L^{D-1}}\int_0^\infty dt\,
t^{D/2-1}\vartheta_3({0,it/\pi})
\exp\left[-\frac{l_p^2}{L^2}t\right]\,,
\end{eqnarray}
where $\vartheta_3$ denotes the Jacobi theta function (see Appendix \ref{AA}).
Integrating the series by term, we find
\begin{eqnarray}
U_D^{(P)}(L)&=&-\frac{\Gamma(D/2)}{2\pi^{D/2}}\sum_{n=-\infty}^{\infty}
\frac{L}{\left(l_p^2+n^2L^2\right)^{D/2}}\nonumber \\
&=&-\frac{\Gamma(D/2)L}{2\pi^{D/2}l_p^D}
-\frac{\Gamma(D/2)}{\pi^{D/2}}\sum_{n=1}^{\infty}
\frac{L}{\left(l_p^2+n^2L^2\right)^{D/2}}\,.
\end{eqnarray}
Since the first term in the last expression is the large $L$ limit of $U_D^{(P)}$,
 we subtract it and define the refined (or ``renormalized'')
effective potential
\begin{equation}
\bar{U}_D^{(P)}(L)=U_D^{(P)}(L)-U_D^{(P)}(\infty)=
-\frac{\Gamma(D/2)}{\pi^{D/2}}\sum_{n=1}^{\infty}
\frac{L}{\left(l_p^2+n^2L^2\right)^{D/2}}\,.
\end{equation}

It should be noticed that,
even if the one-loop vacuum energy density $U_D^{(P)}$ is finite due to UV
completion, the huge contribution of the order of $l_p^{-D}$ should be renormalized
or compensated by some method. This attitude is taken in the subsequent sections. 
We will come back to the discussion in Sec.~\ref{conclusion}.

Now, the known result on standard one-loop scalar energy density for the
Kaluza-Klein theory,
$-\frac{\Gamma(D/2)\zeta_R(D)}{\pi^{D/2}L^{D-1}}$ (where $\zeta_R(z)$ is the
Riemann's zeta function) \cite{CW}, can be obtained by taking the limit
$l_p\rightarrow 0$. 
Expanding $\bar{U}_D^{(P)}$ in terms of a small $l_p/L$, we obtain
\begin{eqnarray}
\bar{U}_D^{(P)}(L)&=&
-\frac{\Gamma(D/2)}{\pi^{D/2}L^{D-1}}\sum_{n=1}^{\infty}
\frac{1}{n^D}
\sum_{k=0}^\infty\frac{(-1)^k\Gamma(k+D/2)}{\Gamma(D/2)k!}
\left(\frac{l_p^2}{n^2L^2}\right)^k\nonumber \\
&=&-\frac{1}{\pi^{D/2}L^{D-1}}
\sum_{k=0}^\infty\frac{(-1)^k\Gamma(k+D/2)\zeta(D-2k)}{k!}
\left(\frac{l_p^2}{L^2}\right)^k
\,.
\end{eqnarray}
On the other hand, we can obtain another expression by series
\begin{equation}
\bar{U}_D^{(P)}(L)=-\frac{\Gamma((D-1)/2)}{2\pi^{(D-1)/2}l_p^{D-1}}-
2\sum_{n=1}^\infty\left(\frac{n}{l_pL}\right)^{(D-1)/2}
K_{(D-1)/2}(2\pi l_pn/L)\,,
\end{equation}
where, $K_\nu(z)$ is the modified Bessel function of the second kind
(see Appendix \ref{AA}),
which shows that
$\bar{U}_D^{(P)}(0)=-\frac{\Gamma((D-1)/2)}{2\pi^{(D-1)/2}l_p^{D-1}}$ is finite.

%%%%%%%%%%%%%%%%%%%%%%%%%%%%%%%%%%%%%%%%%%%%%%%%%%%%%%%%%%%%%%%%%%%%%%%%%%%

Further, we consider a complex massless scalar field $\Phi$ and assume a
general boundary condition
\begin{equation}
\Phi(y+L)=e^{i\delta}\Phi(y)\,.
\label{tbc}
\end{equation}
Then, the vacuum energy density can be obtained as
\begin{equation}
\bar{U}_D^{(P)}(L,\delta)=
-\frac{2\Gamma(D/2)}{\pi^{D/2}}\sum_{n=1}^{\infty}
\frac{L\cos(n\delta)}{\left(l_p^2+n^2L^2\right)^{D/2}}\,.
\end{equation}
Of course, $\bar{U}_D^{(P)}(L,0)$ coincides with $\bar{U}_D^{(P)}(L)$.
Additionally, $\bar{U}_D^{(P)}(L,\delta)$ has a closed form without integrals and
summations provided that $D$ is an even integer. For instance, for $D=4$, one can
find
\begin{equation}
{\textstyle\bar{U}_4^{(P)}(L,\delta)=
\frac{L}{2\pi^2l_p^4}-
\frac{1}{4\pi l_p^3}
\frac{\cosh[l_p(\pi-\delta)/L]+(l_p/L)\delta
\sinh[l_p(\pi-\delta)/L]}{\sinh(l_p\pi/L)}-\frac{1}{4l_p^2L}
\frac{\cosh(l_p\delta/L)}{\sinh^2(l_p\pi/L)}}\,.
\end{equation}

Numeric analyses will be given in a later subsection, after derivation of vacuum
energy in terms of all the three types of Green's functions, where
we compare the result from each scheme of UV completion.

%%%%%%%%%%%%%%%%%%%%%%%%%%%%%%%%%%%%%%%%%%%%%%%%%%%%%%%%%%%%%%%%%%%%%%%%%%%
%%%%%%%%%%%%%%%%%%%%%%%%%%%%%%%%%%%%%%%%%%%%%%%%%%%%%%%%%%%%%%%%%%%%%%%%%%%
\subsection{Vacuum energy from Abel-type Green's function}
%%%%%%%%%%%%%%%%%%%%%%%%%%%%%%%%%%%%%%%%%%%%%%%%%%%%%%%%%%%%%%%%%%%%%%%%%%%
%%%%%%%%%%%%%%%%%%%%%%%%%%%%%%%%%%%%%%%%%%%%%%%%%%%%%%%%%%%%%%%%%%%%%%%%%%%
We define the effective action in $\mathbf{R}^{D-1}\otimes S^1$ from Abel's heat
kernel, that is.
\begin{eqnarray}
W_D^{(A)}&=&\frac{1}{2}L\int d^{D-1}z\,\, \frac{1}{2}\int_0^\infty \frac{ds}{s}
K_D^{(A)}(0,0;s) =\frac{1}{2}L\int d^{D-1}z \int_{l_p^2/2}^\infty \frac{ds}{s}
K_D^{(A)}(0,0;s)
\nonumber \\
&=&\frac{1}{2}L\int d^{D-1}z \int_{l_p^2}^\infty \frac{dT}{\sqrt{T^2-l_p^4}}
K_D(0,0;T)\,.
\end{eqnarray}
It is notable that the factor $1/2$ in front of the integration over $s$ 
after the first equal sign comes from
the exact duality
$s\leftrightarrow l_p^4/(4s)$ in this one-loop integration.
In a similar way to Padmanabhan's case, we define the refined vacuum energy
density and the expression for this vacuum energy is
\begin{eqnarray}
\bar{U}_D^{(A)}(L)&=&-\frac{1}{2\pi^{D/2}L^{D-1}}\int_0^{\frac{L^2}{4l_p^2}} 
dt\frac{t^{D/2-1}}{\sqrt{1-\frac{16l_p^4}{L^4}t^2}}\left(\sum_{n=-\infty}^\infty
e^{-n^2t}-1\right)\nonumber
\\ &=&-\frac{1}{2\pi^{D/2}L^{D-1}}\int_0^{\frac{L^2}{4l_p^2}} 
dt\frac{t^{D/2-1}}{\sqrt{1-\frac{16l_p^4}{L^4}t^2}}\left[ \vartheta_3(0,it/\pi)-1
\right]\,.
\end{eqnarray}

For a complex scalar field with twisted boundary condition (\ref{tbc}),
the vacuum energy density becomes
\begin{eqnarray}
\bar{U}_D^{(A)}(L,\delta)&=&-\frac{1}{\pi^{D/2}L^{D-1}}\int_0^{\frac{L^2}{4l_p^2}} 
dt\frac{t^{D/2-1}}{\sqrt{1-\frac{16l_p^4}{L^4}t^2}}\left(\sum_{n=-\infty}^\infty
e^{-n^2t+in\delta}-1\right)\nonumber
\\ &=&-\frac{1}{\pi^{D/2}L^{D-1}}\int_0^{\frac{L^2}{4l_p^2}} 
dt\frac{t^{D/2-1}}{\sqrt{1-\frac{16l_p^4}{L^4}t^2}}\left[
\vartheta_3(\delta/(2\pi),it/\pi)-1
\right]\,.
\end{eqnarray}

%%%%%%%%%%%%%%%%%%%%%%%%%%%%%%%%%%%%%%%%%%%%%%%%%%%%%%%%%%%%%%%%%%%%%%%%%%%
%%%%%%%%%%%%%%%%%%%%%%%%%%%%%%%%%%%%%%%%%%%%%%%%%%%%%%%%%%%%%%%%%%%%%%%%%%%
\subsection{Vacuum energy from Siegel-type Green's function}
%%%%%%%%%%%%%%%%%%%%%%%%%%%%%%%%%%%%%%%%%%%%%%%%%%%%%%%%%%%%%%%%%%%%%%%%%%%
%%%%%%%%%%%%%%%%%%%%%%%%%%%%%%%%%%%%%%%%%%%%%%%%%%%%%%%%%%%%%%%%%%%%%%%%%%%
The effective action in $\mathbf{R}^{D-1}\otimes S^1$ from Siegel's heat
kernel can be obtained by
\begin{equation}
W_D^{(S)}=\frac{1}{2}L\int d^{D-1}z \int_{0}^\infty \frac{ds}{s}
K_D^{(S)}(0,0;s)=\frac{1}{2}L\int d^{D-1}z \int_{l_p^2}^\infty \frac{ds}{s}
K_D(0,0;s)\,.
\end{equation}
Similarly to the former cases, we define the refined vacuum energy
density and the expression for is is found to be
\begin{eqnarray}
\bar{U}_D^{(S)}(L)&=&-\frac{1}{2\pi^{D/2}L^{D-1}}\int_0^{\frac{L^2}{l_p^2}} 
dt\,t^{D/2-1}\left(\sum_{n=-\infty}^\infty
e^{-n^2t}-1\right)\nonumber
\\ &=&-\frac{1}{2\pi^{D/2}L^{D-1}}\int_0^{\frac{L^2}{l_p^2}} 
dt\,t^{D/2-1}\left[ \vartheta_3(0,it/\pi)-1
\right]\,,
\end{eqnarray}
and the vacuum energy density from the twisted scalar field becomes
\begin{eqnarray}
\bar{U}_D^{(S)}(L,\delta)&=&-\frac{1}{\pi^{D/2}L^{D-1}}\int_0^{\frac{L^2}{l_p^2}} 
dt\,t^{D/2-1}\left(\sum_{n=-\infty}^\infty
e^{-n^2t+in\delta}-1\right)\nonumber
\\ &=&-\frac{1}{\pi^{D/2}L^{D-1}}\int_0^{\frac{L^2}{l_p^2}} 
dt\,t^{D/2-1}\left[ \vartheta_3(\delta/(2\pi),it/\pi)-1
\right]\,.
\end{eqnarray}

%%%%%%%%%%%%%%%%%%%%%%%%%%%%%%%%%%%%%%%%%%%%%%%%%%%%%%%%%%%%%%%%%%%%%%%%%%%
%%%%%%%%%%%%%%%%%%%%%%%%%%%%%%%%%%%%%%%%%%%%%%%%%%%%%%%%%%%%%%%%%%%%%%%%%%%
\subsection{Numerical comparison of three vacuum energy densities}
%%%%%%%%%%%%%%%%%%%%%%%%%%%%%%%%%%%%%%%%%%%%%%%%%%%%%%%%%%%%%%%%%%%%%%%%%%%
%%%%%%%%%%%%%%%%%%%%%%%%%%%%%%%%%%%%%%%%%%%%%%%%%%%%%%%%%%%%%%%%%%%%%%%%%%%
In this subsection, we show the numerical results for the three types of vacuum
energy in the case of $D=5$.

\begin{figure}[ht]
\centering
\includegraphics%[width=5cm]
{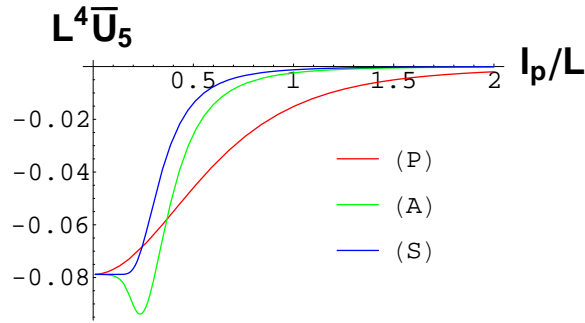}
\caption{The vacuum energy densities regulated as $L^4\bar{U}_5$ are plotted
against the Planck length $l_p$ divided by the circumference $L$ of $S^1$.
The red curve indicates the Padmanabhan-type, the green curve indicates the
Abel-type, and the blue curve indicates the Siegel-type.}
\label{fig1}
\end{figure}

Fig.~\ref{fig1} shows the dependence of the vacuum energies on the magnitude of
the Planck length, $l_p/L$. The value of $L^4\bar{U}_5$ is
$-\frac{3\zeta(5)}{4\pi^2}=-0.0787\cdots$ in the limit of $l_p\rightarrow 0$.
Each dependence on $l_p/L$ is different. The value of $L^4\bar{U}_5$ of Abel-type
takes a minimum at a finite $l_p$. The value of $L^4\bar{U}_5$ of Siegel-type
varies moderately near $l_p/L\ll 1$. Note that, as we simply take a common scale
$l_p$ for three cases, the comparison in values at the same $l_p/L$ in the figure
has only qualitative meanings.

%%%%%%%%%%%%%%%%%%%%%%%%%%%
% 1
%%%%%%%%%%%%%%%%%%%%%%%%%%%
%\begin{wrapfigure}{r}{5cm}
%\end{wrapfigure}
%%%%%%%%%%%%%%%%%%%%%%%%%%%

\begin{figure}[ht]
\centering
\includegraphics%[width=5cm]
{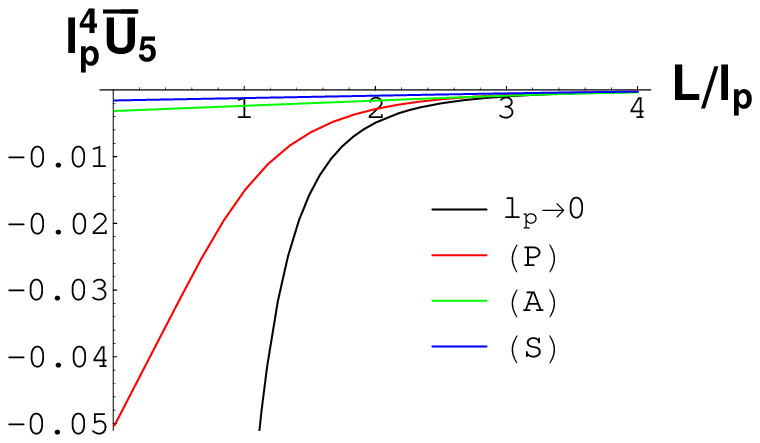}
\includegraphics%[width=5cm]
{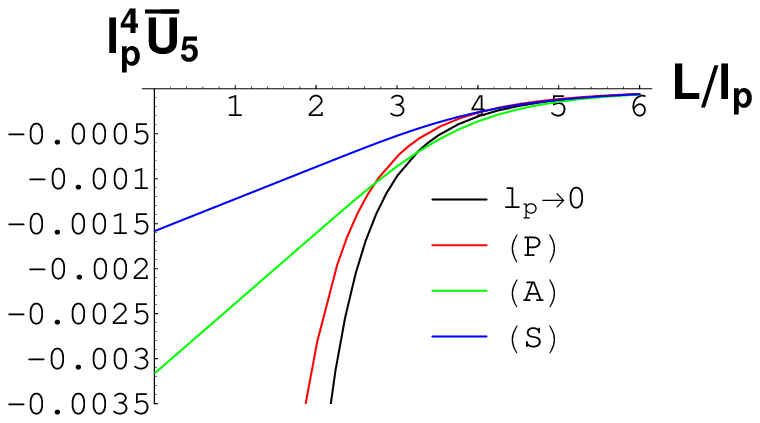}
\\
(a) \hspace{8cm} (b)
\caption{The vacuum energy densities $l_p^4\bar{U}_5$ are plotted
against the circumference $L$ of $S^1$ divided by the Planck length $l_p$.
The black curve indicates the usual case, i.e., the case that the Planck length is
set to zero, the red curve indicates the Padmanabhan-type, the green curve
indicates the Abel-type, and the blue curve indicates the Siegel-type.
The left plot (a) shows the range $L/l_p<4$, while the right plot (b) shows the
range $L/l_p<6$ with an enlarged vertical axis.}
\label{fig2}
\end{figure}

Fig.~\ref{fig2} shows the vacuum energy densities $l_p^4\bar{U}_5$ as functions of
$L/l_p$. All the vacuum energy densities calculated from three UV improved Green's
functions have finite values at $L=0$. All of the absolute value of vacuum energy
densities are monotonically decreasing as functions of $L$. The deviation from the
standard case without the fundamental scale becomes large at
$L/l_p< 1$. The comparison of absolute values has little
meaning, because the Planck length for each three scheme is basically defined as
an individual value. For large
$L/l_p$, however, all the values for the energy densities are indistinguishable
as expected.

%%%%%%%%%%%%%%%%%%%%%%%%%%%
% 2
%%%%%%%%%%%%%%%%%%%%%%%%%%%
%\begin{wrapfigure}{r}{5cm}
%\end{wrapfigure}
%%%%%%%%%%%%%%%%%%%%%%%%%%%

\begin{figure}[ht]
\centering
\includegraphics[width=5cm]{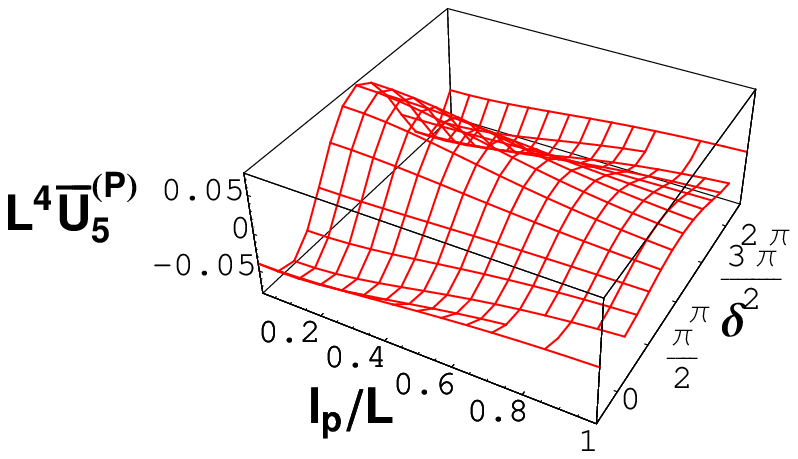}\quad
\includegraphics[width=5cm]{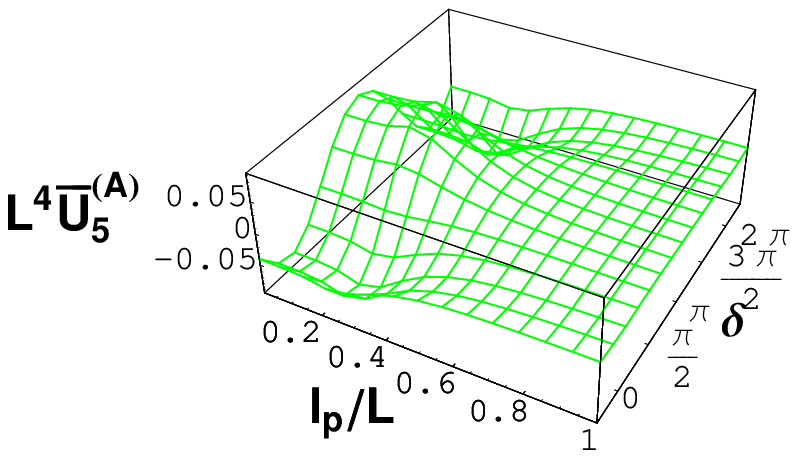}\quad
\includegraphics[width=5cm]{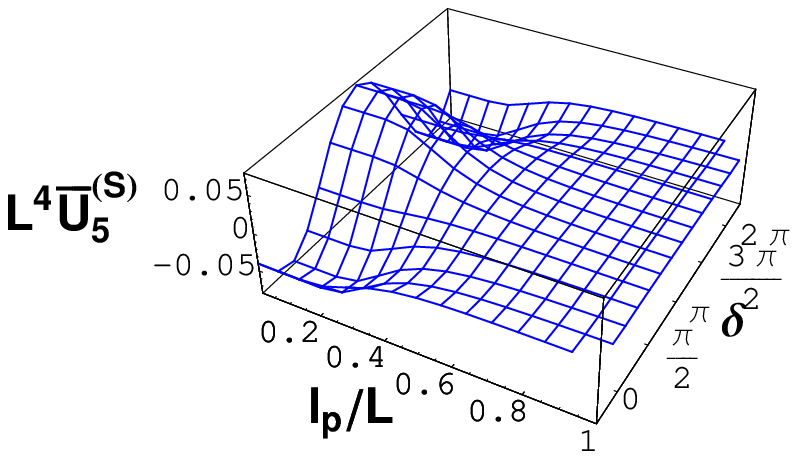}
\\
(P) \hspace{5cm} (A) \hspace{5cm} (S)
\caption{Plots of the vacuum energy densities of a complex scalar field with the twisted boundary
condition for (P) Padmanabhan-type,
(A) Abel-type,  and (S) Siegel-type.}
\label{fig3}
\end{figure}

The vacuum energy densities of a complex scalar field with the twisted boundary
condition $\bar{U}_5(L,\delta)$ are exhibited in Fig.~\ref{fig3}.
If we assume that $L$ is kept at a finite value, a finite value of the Planck
length makes the effective potential with respect to $\delta$ flat in all the
three cases. The twisted parameter can be regarded as a dynamical variable, which
comes from the vacuum gauge field on $S^1$ \cite{Hosotani}.
Furthermore, the possibility of the identification of such a degree of freedom
as an inflaton has been proposed by several authors \cite{ACCR,IKLM}.
Because the flat potential would be suitable for such inflationary scenarios,
it can be said that the effect of UV cutoff may also be relevant to the
cosmological dynamics.

%%%%%%%%%%%%%%%%%%%%%%%%%%%
% 3
%%%%%%%%%%%%%%%%%%%%%%%%%%%
%\begin{wrapfigure}{r}{5cm}
%\end{wrapfigure}
%%%%%%%%%%%%%%%%%%%%%%%%%%%

%%%%%%%%%%%%%%%%%%%%%%%%%%%%%%%%%%%%%%%%%%%%%%%%%%%%%%%%%%%%%%%%%%%%%%%%%%%
%%%%%%%%%%%%%%%%%%%%%%%%%%%%%%%%%%%%%%%%%%%%%%%%%%%%%%%%%%%%%%%%%%%%%%%%%%%
%%%%%%%%%%%%%%%%%%%%%%%%%%%%%%%%%%%%%%%%%%%%%%%%%%%%%%%%%%%%%%%%%%%%%%%%%%%
\section{UV improved quantum stress tensors in a conical space}
\label{sec4}
%%%%%%%%%%%%%%%%%%%%%%%%%%%%%%%%%%%%%%%%%%%%%%%%%%%%%%%%%%%%%%%%%%%%%%%%%%%
%%%%%%%%%%%%%%%%%%%%%%%%%%%%%%%%%%%%%%%%%%%%%%%%%%%%%%%%%%%%%%%%%%%%%%%%%%%
%%%%%%%%%%%%%%%%%%%%%%%%%%%%%%%%%%%%%%%%%%%%%%%%%%%%%%%%%%%%%%%%%%%%%%%%%%%

Let us turn to another non-trivial space, a conical space or space with a conical
defect.
We take the coordinates for a conical space as
\begin{equation}
ds^2=\sum_{i=1}^{D-2}(dz^i)^2+dr^2+\frac{r^2}{\nu^2}d\theta^2\,,
\end{equation}
where $\nu$ is a constant greater than unity. This metric is equivalent to
\begin{equation}
ds^2=\sum_{i=1}^{D-2}(dz^i)^2+dr^2+r^2 d\tilde{\theta}^2\,,
\label{flat}
\end{equation}
where the range of $\tilde{\theta}$ is $0<\tilde{\theta}\le 2\pi/\nu$.
This metric adequately describes a locally flat Euclidean space except for
the coordinate origin if $\nu\ne 1$. A space with a deficit angle is often
employed as  a model space around a mathematically idealized straight cosmic string
\cite{Vilenkin}.

The standard heat kernel in a conical space without fundamental length is known
\cite{CKV,Moreira} and presented in the form 
in terms of $x=(r,\theta, z^i)$ and $x'=(r',\theta', z'^i)$,
\begin{eqnarray}
K_{D,\nu}(x,x';s)&=&\frac{\nu}{(4\pi
s)^{D/2}}\sum_{m=-\infty}^{\infty}e^{im\varphi}
I_{\nu|m|}\left(\frac{rr'}{2s}\right)\exp\left[-\frac{r^2+{r'}^2+\zeta^2}{4s}
%-M^2s
\right]\nonumber \\
&\equiv& K_{D,\nu}(r,r',\varphi,\zeta;s)\,,
\label{conK}
\end{eqnarray}
where $\varphi=\theta-\theta'$ and $\zeta=|z-z'|$.
By using the formula
\begin{eqnarray}
& &\sum_{m=-\infty}^\infty e^{im\varphi}
I_{\nu|m|}(z)=\frac{1}{\nu}\exp[{z\cos(\varphi/\nu)}]\nonumber \\
& &\qquad+\frac{1}{2\pi}\int_{0}^\infty
e^{-z\cosh v}\left[\frac{\sin(\varphi-\nu\pi)}{\cosh\nu
v-\cos(\varphi-\nu\pi)}-\frac{\sin(\varphi+\nu\pi)}{\cosh\nu
v-\cos(\varphi+\nu\pi)}\right]dv\,,
\end{eqnarray}
which can be derived from the integral form of the modified Bessel function 
\begin{equation}
I_\mu(z)=\frac{1}{2\pi}\int_{-\pi}^\pi e^{z\cos\theta}\cos\mu\theta\,d\theta
-\frac{\sin\mu\pi}{\pi}\int_0^\infty
e^{-z\cosh v-\mu v}\,dv\,,
\end{equation}
the heat kernel $K_{D,\nu}(r,r',\varphi,\zeta; s)$ can be recast in the form
\begin{eqnarray}
& &K_{D,\nu}(r,r',\varphi,\zeta;s)
={\textstyle\frac{1}{(4\pi
s)^{D/2}}}\exp\left[{\textstyle
-\frac{r^2+{r'}^2-2rr'\cos\tilde{\varphi}+\zeta^2}{4s}
%-M^2s
}\right]\nonumber \\
& &\,+{\textstyle\frac{e^{-\frac{r^2+{r'}^2+\zeta^2}{4s}
%-M^2s
}}{2\pi(4\pi
s)^{D/2}}}\int_{0}^\infty
e^{-\frac{rr'}{2s}\cosh v}\left[{\textstyle
\frac{\nu\sin\nu(\tilde{\varphi}-\pi)}{\cosh\nu
v-\cos\nu(\tilde{\varphi}-\pi)}-\frac{\nu\sin\nu(\tilde{\varphi}+\pi)}{\cosh\nu
v-\cos\nu(\tilde{\varphi}+\pi)}}\right]dv\,,
\label{eq00}
\end{eqnarray}
where $\tilde{\varphi}=\varphi/\nu$.

The first term in the right hand side of (\ref{eq00}) coincides with the heat
kernel $K_D(x,x';s)$ in the locally flat space (\ref{flat}).
One can also find that the expression in the brackets in the second term in the
right hand side of (\ref{eq00})  vanishes when $\nu=1$.
Therefore, it is convenient to define the refined heat kernel
$\bar{K}_{D,\nu}(x,x';s)\equiv{K}_{D,\nu}(x,x';s)-{K}_{D}(x,x';s)$, i.e.,
\begin{eqnarray}
\bar{K}_{D,\nu}(x,x';s)&=&\bar{K}_{D,\nu}(r,r',\varphi,\zeta;s)\nonumber \\
&\equiv&{\textstyle
\frac{e^{-\frac{r^2+{r'}^2+\zeta^2}{4s}
}}{2\pi(4\pi
s)^{D/2}}}\int_{0}^\infty
e^{-\frac{rr'}{2s}\cosh v}\left[{\textstyle
\frac{\nu\sin\nu(\tilde{\varphi}-\pi)}{\cosh\nu
v-\cos\nu(\tilde{\varphi}-\pi)}-\frac{\nu\sin\nu(\tilde{\varphi}+\pi)}{\cosh\nu
v-\cos\nu(\tilde{\varphi}+\pi)}}\right]dv\,.
\label{sk}
\end{eqnarray}

Now, we introduce three types of refined heat kernels in the conical space,
as the previously-used way.
They are:
\begin{eqnarray}
\bar{K}_{D,\nu}^{(P)}(r,r',\varphi,\zeta;s)&=&
\bar{K}_{D,\nu}(r,r',\varphi,{\textstyle\sqrt{
\zeta^2+l_p^2}};s)\\
\bar{K}_{D,\nu}^{(A)}(r,r',\varphi,\zeta;s)&=&
\bar{K}_{D,\nu}(r,r',\varphi,\zeta;{\textstyle s+\frac{l_p^4}{4s}})\\
\bar{K}_{D,\nu}^{(S)}(r,r',\varphi,\zeta;s)&=&
\bar{K}_{D,\nu}(r,r',\varphi,\zeta;s+l_p^2)\,.
\end{eqnarray}

The expectation value of the quantum stress tensor operator for a massless scalar
field is given by the limit, as in the case with the standard Green's function
\cite{CKV,Moreira,Linet,FS,Smith,SH}
\footnote{In the present paper, we disregard the possible modification on the
stress tensor operator as well as the reaction to the background metric.}
\begin{equation}
\langle T^\rho_{\sigma}\rangle^{(*)}=\lim_{x'\rightarrow
x}D^\rho_{\sigma}\bar{G}_{D,\nu}^{(*)}(x,x')\qquad
({\textstyle(*)=(P), (A), \mbox{~and~} (S)})\,,
\label{qst}
\end{equation}
where $\bar{G}_{D,\nu}^{(*)}(x,x')=\int_0^\infty
\bar{K}_{D,\nu}^{(*)}(x,x';s) ds$ and $D^\rho_\sigma$ is the second order
differential operator
\begin{equation}
D^\rho_{\sigma}
\equiv(1-2\xi)\nabla^\rho\nabla_{\sigma'}-\left(\frac{1}{2}-2\xi\right)
\delta^\rho_\sigma\nabla^\lambda\nabla_{\lambda'}-2\xi(\nabla^\rho\nabla_\sigma
-\delta^\rho_\sigma\nabla^\lambda\nabla_\lambda)\,.
\label{qstd}
\end{equation}
Here, $\nabla_\sigma$ stands for a covariant derivative and $\xi$ is the coupling
between the scalar field and the Ricci curvature
$R$, which modified the scalar Laplacian $-\Delta\rightarrow -\Delta+\xi R$. In the
conical space presently considered, the curvature is zero everywhere outside the
conical singularity and the heat kernels and Green's functions are unchanged.
Thus, the aforementioned relation
$\bar{G}_{D,\nu}^{(*)}(x,x')=\int_0^\infty
\bar{K}_{D,\nu}^{(*)}(x,x';s) ds$ is held even in the case of $\xi\ne 0$.

Hereafter, we concentrate ourselves on the case with $D=4$, unless especially
mentioned on $D$.
As in the preceding section, the numerical estimation will be exhibited all at
once in the last subsection.

%%%%%%%%%%%%%%%%%%%%%%%%%%%%%%%%%%%%%%%%%%%%%%%%%%%%%%%%%%%%%%%%%%%%%%%%%%%
%%%%%%%%%%%%%%%%%%%%%%%%%%%%%%%%%%%%%%%%%%%%%%%%%%%%%%%%%%%%%%%%%%%%%%%%%%%
\subsection{The refined Green's function and quantum stress tensor}
%%%%%%%%%%%%%%%%%%%%%%%%%%%%%%%%%%%%%%%%%%%%%%%%%%%%%%%%%%%%%%%%%%%%%%%%%%%
%%%%%%%%%%%%%%%%%%%%%%%%%%%%%%%%%%%%%%%%%%%%%%%%%%%%%%%%%%%%%%%%%%%%%%%%%%%

It is known that the standard, full (not refined) Green's function without a cutoff
scale in a conical space can be written in a simple closed form for $D=4$
\cite{Smith}. Thus, the Padmanabhan-type Green's function is easily obtained by
the replacement
$\zeta\rightarrow {\sqrt{\zeta^2+l_p^2}}$ as
\begin{equation}
G_{4,\nu}^{(P)}(x,x')=\frac{1}{8\pi^2
r r'\sinh
u}\frac{\nu \sinh\nu u}{\cosh\nu u-\cos\nu\tilde{\varphi}}\,,
\label{cf}
\end{equation}
where $\cosh u=\frac{r^2+r'^2+\zeta^2+l_p^2}{2 r r'}$ or
$\sinh \frac{u}{2}=\sqrt{\frac{(r-r')^2+\zeta^2+l_p^2}{4 r r'}}$.
As for the Padmanabhan-type vacuum averages, the use of this expression is
easily handled.

First of all, we consider the vacuum expectation value of $\phi^2$ in the conical
space
\cite{Smith}. In the present case, this is given by
\begin{equation}
\langle\phi^2\rangle^{(P)}(x)=\bar{G}_{4,\nu~}^{(P)}(x,x)\,,
\end{equation}
where
\begin{equation}
\bar{G}_{4,\nu}^{(P)}(x,x')=G_{4,\nu}^{(P)}(x,x')-G_{4,1}^{(P)}(x,x')\,.
\end{equation}
A straightforward calculation with the expression (\ref{cf}) yields
\begin{equation}
\langle\phi^2\rangle^{(P)}(r)=\frac{1}{4\pi^2l_p^2}\left(f_1-1
\right)\,,
\label{vf1}
\end{equation}
where
\begin{equation}
f_1\equiv\frac{\nu l_p\coth\left(\nu\sinh^{-1}\frac{l_p}{2r}
\right)}{\sqrt{l_p^2+4r^2}}\,.
\label{vf2}
\end{equation}
It is worth noting that $f_1\equiv 1$ for $\nu=1$, while $\lim_{l_p/r\rightarrow
0}f_1=1$ and $\lim_{l_p/r\rightarrow \infty}f_1=\nu$ for $\nu>1$.

For small $l_p^2/r^2$, we find
\begin{equation}
\langle\phi^2\rangle^{(P)}(r)=\frac{\nu^2-1}{48\pi^2r^2}\left(1-
\frac{(\nu^2+11)l_p^2}{60r^2}
+O(l_p^4/r^4)\right)\,.
\end{equation}
Therefore, the effect of the Planck length vanishes far from the conical
singularity sitting on the origin. Of course, the limit of $l_p=0$ gives the
standard result in a conical space \cite{Smith}.

%On the other hand, on the conical singularity,
%the expectation value is finite:
%\begin{equation}
%\lim_{r\rightarrow
%0}\langle\phi^2\rangle^{(P)}(r)=\frac{\nu-1}{4\pi^2l_p^2}\,.
%\end{equation}

Now, we consider the quantum stress tensor of Padmanabhan-type in the conical
space. Using the formula (\ref{qst}), we obtain%
\footnote{Bear in mind that $\nabla_{\tilde{\theta}}\nabla_{\tilde{\theta}}=
\partial_{\tilde{\theta}}^2+r\partial_r$.}
\begin{eqnarray}
\langle
T^r_r\rangle^{(P)}&=&\frac{1}{\pi^2l_p^4}\left[
-\frac{r^2}{l_p^2+4r^2}f_3+\frac{l_p^4-8l_p^2r^2-12r^4}{6(l_p^2+4r^2)^2}f_2
-\frac{2l_p^4+5l_p^2r^2+6r^4}{3(l_p^2+4r^2)^2}f_1+\frac{1}{2}\right]\nonumber \\
& &+\frac{1}{\pi^2l_p^2}\left(\xi-\frac{1}{6}\right)\frac{f_2-f_1}{l_p^2+4r^2}
\,,\\
\langle
T^{\tilde{\theta}}_{\tilde{\theta}}\rangle^{(P)}&=&\frac{1}{\pi^2l_p^4}\left[
\frac{l_p^2+3r^2}{3(l_p^2+4r^2)}f_3-\frac{l_p^4+16l_p^2r^2+36r^4}{6(l_p^2+4r^2)^2}f_2
-\frac{2l_p^4+11l_p^2r^2+18r^4}{3(l_p^2+4r^2)^2}f_1+\frac{1}{2}\right]\nonumber \\
&
&+\frac{1}{\pi^2l_p^2}\left(\xi-\frac{1}{6}\right)\left[
\frac{2}{l_p^2+4r^2}f_3-\frac{l_p^2+16r^2}{(l_p^2+4r^2)^2}f_2
-\frac{l_p^2-8r^2}{(l_p^2+4r^2)^2}f_1
\right]
\,,\\
\langle
T^{z}_{z}\rangle^{(P)}&\equiv&
\langle
T^{z_1}_{z_1}\rangle^{(P)}=\langle
T^{z_2}_{z_2}\rangle^{(P)}\nonumber \\
&=&\frac{1}{\pi^2l_p^4}\left[
-\frac{l_p^2+6r^2}{6(l_p^2+4r^2)}f_3-\frac{r^2(l_p^2+2r^2)}{(l_p^2+4r^2)^2}f_2
-\frac{l_p^4+4l_p^2r^2+6r^4}{3(l_p^2+4r^2)^2}f_1+\frac{1}{2}\right]\nonumber \\
&
&+\frac{1}{\pi^2l_p^2}\left(\xi-\frac{1}{6}\right)\left[
\frac{2}{l_p^2+4r^2}f_3-\frac{12r^2}{(l_p^2+4r^2)^2}f_2
-\frac{2(l_p^2-2r^2)}{(l_p^2+4r^2)^2}f_1
\right]
\,,
\end{eqnarray}
where
\begin{equation}
f_2\equiv\frac{\nu^2
l_p^2}{4r^2\sinh^2\left(\nu\sinh^{-1}\frac{l_p}{2r}\right)}\,,\qquad
f_3\equiv\frac{\nu^3 l_p^3\coth\left(\nu\sinh^{-1}\frac{l_p}{2r}
\right)}{4r^2\sqrt{l_p^2+4r^2}\sinh^2\left(\nu\sinh^{-1}\frac{l_p}{2r}\right)}\,.
\end{equation}
It should be noticed that $f_2=f_3\equiv 1$ for $\nu=1$, while
$\lim_{l_p/r\rightarrow 0}f_2=\lim_{l_p/r\rightarrow 0}f_3=1$ and
$\lim_{l_p/r\rightarrow \infty}f_3=\lim_{l_p/r\rightarrow \infty}f_3=0$ for
$\nu>1$.

For small $l_p^2/r^2$, they reveal
\begin{eqnarray}
\langle T^r_r\rangle^{(P)}&=&\frac{\nu^4-1}{1440\pi^2r^4}
-\frac{(\nu^2-1)(\nu^4+8\nu^2+57)l_p^2}{15120\pi^2r^6}\nonumber \\
& &+\left(\xi-\frac{1}{6}\right)
\left[-\frac{\nu^2-1}{24\pi^2r^4}
+\frac{(\nu^2-1)(\nu^2+11)l_p^2}{720\pi^2r^6}\right]+O(l_p^4/r^8)\,,\\
\langle
T^{\tilde{\theta}}_{\tilde{\theta}}\rangle^{(P)}&=&-\frac{\nu^4-1}{480\pi^2r^4}
+\frac{(\nu^2-1)(4\nu^2-1)(\nu^2+3)l_p^2}{30240\pi^2r^6}\nonumber \\ &
&+\left(\xi-\frac{1}{6}\right)
\left[\frac{\nu^2-1}{8\pi^2r^4}
-\frac{(\nu^2-1)(\nu^2+11)l_p^2}{144\pi^2r^6}\right]+O(l_p^4/r^8)\,,\\
\langle T^{z}_{z}\rangle^{(P)}&=&\frac{\nu^4-1}{1440\pi^2r^4}
-\frac{(\nu^2-1)(2\nu^4+9\nu^2+37)l_p^2}{30240\pi^2r^6}\nonumber \\
& &+\left(\xi-\frac{1}{6}\right)
\left[\frac{\nu^2-1}{12\pi^2r^4}
-\frac{(\nu^2-1)(\nu^2+11)l_p^2}{180\pi^2r^6}\right]+O(l_p^4/r^8)\,.
\end{eqnarray}
Again, the limit of $l_p=0$ gives the
standard results in a conical space \cite{Smith}.

%%%%%%%%%%%

We find that $\nabla_\mu \langle T^{\mu\nu}\rangle^{(P)}\ne 0$ for
finite $l_p$. That is
\begin{eqnarray}
\nabla_\lambda\langle
T^{\lambda}_{r}\rangle^{(P)}&=&
\partial_r\langle
T^r_r\rangle^{(P)}+{\textstyle\frac{1}{r}}(\langle T^r_r\rangle^{(P)}-\langle
T^{\tilde{\theta}}_{\tilde{\theta}}\rangle^{(P)})\nonumber
\\ &=&
\frac{(\nu^2-1)(2\nu^4+23\nu^2+191)l_p^2}{10080\pi^2r^7}+O(l_p^4/r^9)\,.
\end{eqnarray}
This is obvious, because Padmanabhan-type Green's function does not satisfy
$-\Delta_x{G}^{(P)}_D(x,x')=\frac{1}{\sqrt{g}}\delta^D(x,x')$ for finite $l_p$.
By the way, Abel-type and Siegel-type Green's functions also do not satisfy
the relation. The conservation of the stress tensor will be established in the
region $r\gg l_p$.

The trace of the quantum stress tensor is found to be
\begin{eqnarray}
\langle
T^{\lambda}_{\lambda}\rangle^{(P)}&=&
-\frac{(\nu^2-1)(2\nu^4+23\nu^2+191)l_p^2}{30240\pi^2r^6}\nonumber \\ &
&+\left(\xi-\frac{1}{6}\right)
\left[\frac{\nu^2-1}{4\pi^2r^4}
-\frac{(\nu^2-1)(\nu^2+11)l_p^2}{60\pi^2r^6}\right]+O(l_p^4/r^8)\,,
\end{eqnarray}
and it is not zero for finite $l_p$ even if $\xi=1/6$ (the conformal coupling),
but tends to zero as $r/l_p$ increases.

%Numerical values of the above results on Padmanabhan's type will be given
%together with those of other types in the next subsection in the present section.

%%%%%%%%%%%%%%%%%%%%%%%%%%%%%%%%%%%%%%%%%%%%%%%%%%%%%%%%%%%%%%%%%%%%%%%%%%%

For calculation of Abel-type Green's function, we use the refined heat kernel 
mentioned at the beginning of the present section. We find
\begin{eqnarray}
& &\bar{G}_{4,\nu}^{(A)}(x,x')=\int_0^\infty
\bar{K}_{4,\nu}^{(A)}(x,x';s)ds
\nonumber \\
&
&=
\int_{l_p^2}^{\infty}{\textstyle\frac{TdT}{\sqrt{T^2-l_p^4}}
\frac{e^{-\frac{r^2+{r'}^2+\zeta^2}{4T}
}}{2\pi(4\pi T)^{2}}}\int_{0}^\infty
e^{-\frac{rr'}{2T}\cosh v}
{\textstyle
\left[\frac{\nu\sin\nu(\tilde{\varphi}-\pi)}{\cosh\nu
v-\cos\nu(\tilde{\varphi}-\pi)}-\frac{\nu\sin\nu(\tilde{\varphi}+\pi)}{\cosh\nu
v-\cos\nu(\tilde{\varphi}+\pi)}\right]
}
dv
\nonumber \\
&
&=\int_{0}^{1}{\textstyle\frac{dt}{l_p^2\sqrt{1-t^2}}
\frac{e^{-\frac{r^2+{r'}^2+\zeta^2}{4l_p^2}t
}}{2\pi(4\pi)^{2}}}\int_{0}^\infty
e^{-\frac{rr'}{2l_p^2}t\cosh
v}\left[{\textstyle
\frac{\nu\sin\nu(\tilde{\varphi}-\pi)}{\cosh\nu
v-\cos\nu(\tilde{\varphi}-\pi)}-\frac{\nu\sin\nu(\tilde{\varphi}+\pi)}{\cosh\nu
v-\cos\nu(\tilde{\varphi}+\pi)}}
\right]dv\nonumber \\
&
&=
{\textstyle\frac{1}{64\pi^{2}l_p^2}}\int_{0}^\infty
\left[I_0(
{\scriptstyle
\frac{r^2+{r'}^2+2rr'\cosh
v+\zeta^2}{4l_p^2}
}
)-\mathbf{L}_0(
{\scriptstyle
\frac{r^2+{r'}^2+2rr'\cosh
v+\zeta^2}{4l_p^2}
}
)\right]
{\textstyle
\left[\frac{\nu\sin\nu(\tilde{\varphi}-\pi)}{\cosh\nu
v-\cos\nu(\tilde{\varphi}-\pi)}-\frac{\nu\sin\nu(\tilde{\varphi}+\pi)}{\cosh\nu
v-\cos\nu(\tilde{\varphi}+\pi)}\right]
}
dv\,.\nonumber \\
\end{eqnarray}

%%%%%%%%%%%%%%%%%%%%%%%%%%%%%%%%%%%%%%%%%%%%%%%%%%%%%%%%%%%%%%%%%%%%%%%%%%%

For Siegel-type Green's function, we again use the refined heat kernel
and obtain refined Green's function. That is,
\begin{eqnarray}
& &\bar{G}_{4,\nu}^{(S)}(x,x')=\int_0^\infty
\bar{K}_{4,\nu}^{(S)}(x,x') ds
\nonumber \\
&
&=\int_{l_p^2}^{\infty}ds
\frac{e^{-\frac{r^2+{r'}^2+\zeta^2}{4s}
}}{2\pi(4\pi s)^{2}}\int_{0}^\infty
e^{-\frac{rr'}{2s}\cosh v}\left[{\textstyle
\frac{\nu\sin\nu(\tilde{\varphi}-\pi)}{\cosh\nu
v-\cos\nu(\tilde{\varphi}-\pi)}-\frac{\nu\sin\nu(\tilde{\varphi}+\pi)}{\cosh\nu
v-\cos\nu(\tilde{\varphi}+\pi)}}
\right]dv\nonumber \\
&
&=
\frac{1}{8\pi^{3}}\int_{0}^\infty
{\textstyle\frac{1-e^{-\frac{r^2+{r'}^2+2rr'\cosh
v+\zeta^2}{4l_p^2}
}}{r^2+{r'}^2+2rr'\cosh
v+\zeta^2}}
\left[{\textstyle
\frac{\nu\sin\nu(\tilde{\varphi}-\pi)}{\cosh\nu
v-\cos\nu(\tilde{\varphi}-\pi)}-\frac{\nu\sin\nu(\tilde{\varphi}+\pi)}{\cosh\nu
v-\cos\nu(\tilde{\varphi}+\pi)}}
\right]dv\,.
\end{eqnarray}

From these refined Green's functions, we can enumerate the vacuum fluctuation
$\langle\phi^2\rangle$ and the quantum stress tensor in an analogous way. 
Numerical results obtained from the Green's functions will be given in the next
subsection.

%%%%%%%%%%%%%%%%%%%%%%%%%%%%%%%%%%%%%%%%%%%%%%%%%%%%%%%%%%%%%%%%%%%%%%%%%%%
%%%%%%%%%%%%%%%%%%%%%%%%%%%%%%%%%%%%%%%%%%%%%%%%%%%%%%%%%%%%%%%%%%%%%%%%%%%
\subsection{Numerical comparison of dependence of three expectation values for
$\phi^2$ and three stress tensors on the Planck length}
%%%%%%%%%%%%%%%%%%%%%%%%%%%%%%%%%%%%%%%%%%%%%%%%%%%%%%%%%%%%%%%%%%%%%%%%%%%
%%%%%%%%%%%%%%%%%%%%%%%%%%%%%%%%%%%%%%%%%%%%%%%%%%%%%%%%%%%%%%%%%%%%%%%%%%%

\begin{figure}[ht]
\centering
\includegraphics[width=5cm]{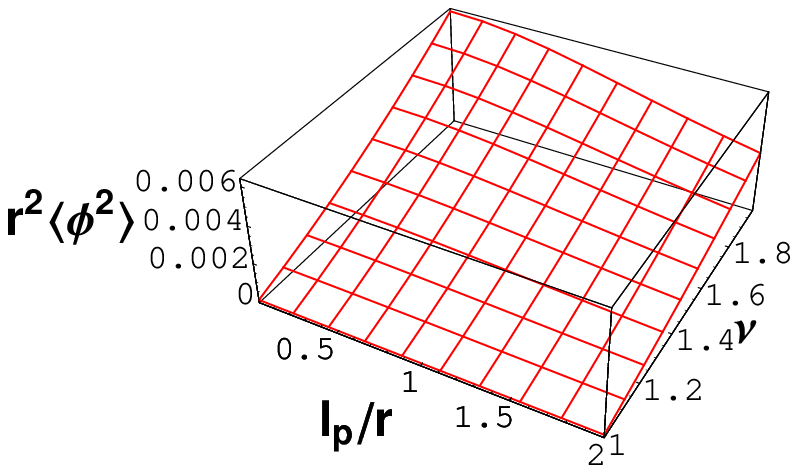}\quad
\includegraphics[width=5cm]{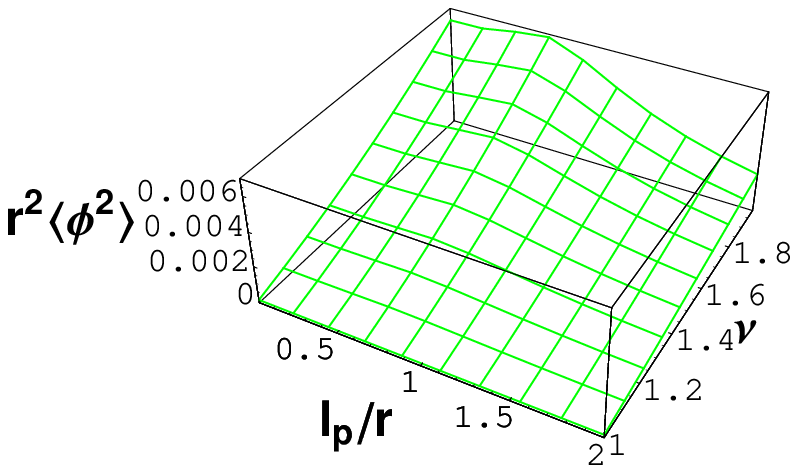}\quad
\includegraphics[width=5cm]{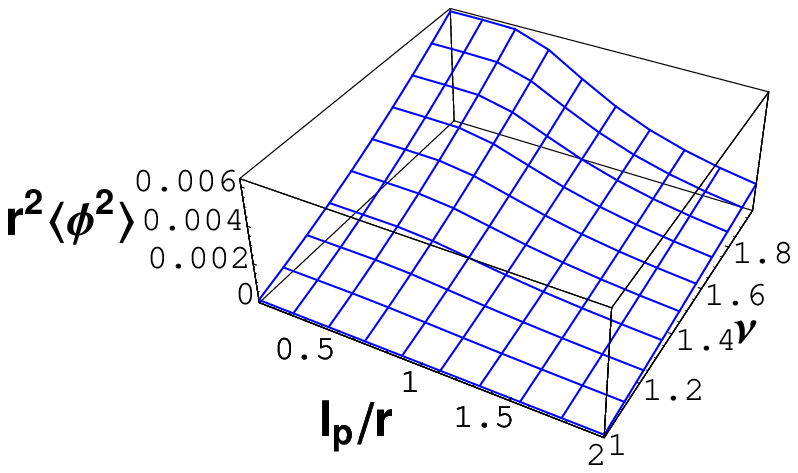}
\\
(P) \hspace{5cm} (A) \hspace{5cm} (S)
\caption{$r^2\langle\phi^2\rangle$ is plotted against $l_p/r$ and $\nu$, for (P)
Padmanabhan-type, (A) Abel-type,  and (S) Siegel-type.}
\label{fig4}
\end{figure}

The expectation value of $\phi^2$ is given by
\begin{equation}
\langle\phi^2\rangle^{(*)}(x)=\bar{G}_{4,\nu~}^{(*)}(x,x)
\qquad ({\textstyle(*)=(P), (A), \mbox{~and~} (S)})\,,
\end{equation}
for three types of Green's functions.
Fig.~\ref{fig4} shows the values of $r^2\langle\phi^2\rangle$ as functions of
$l_p/r$ and $\nu$ for three types.
Except for the slight rise found in the Able's type,
the expectation values of $\phi^2$ decrease (and remain positive)
for larger values of the Planck length at fixed $r$, and they increase as $\nu$
increases as in the standard case with no cutoff scale \cite{Smith}. Of course,
we find that 
$\lim_{l_p/r\rightarrow}r^2\langle\phi^2\rangle=\frac{\nu^2-1}{48\pi^2}\approx
0.00211086\times (\nu^2-1)$ for all the types.

%%%%%%%%%%%%%%%%%%%%%%%%%%%
% 4
%%%%%%%%%%%%%%%%%%%%%%%%%%%
%\begin{wrapfigure}{r}{5cm}
%\end{wrapfigure}
%%%%%%%%%%%%%%%%%%%%%%%%%%%

\begin{figure}[ht]
\centering
\includegraphics[width=5cm]{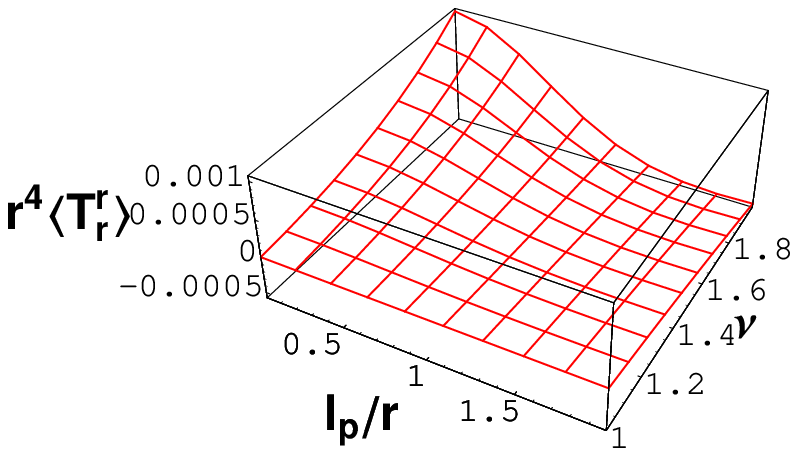}\quad
\includegraphics[width=5cm]{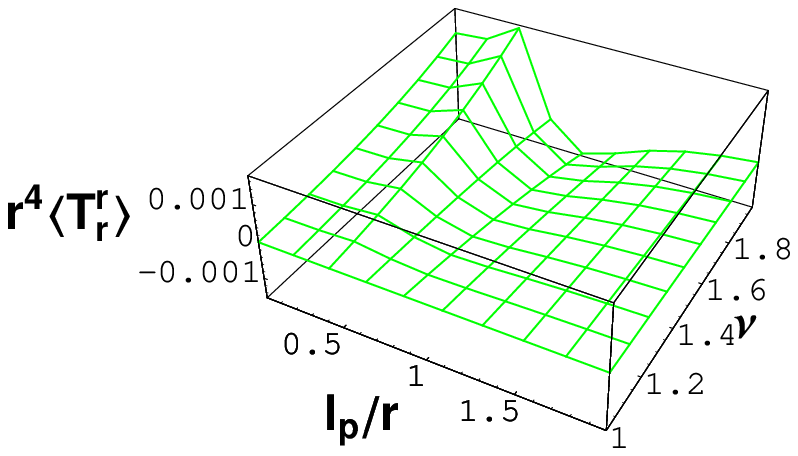}\quad
\includegraphics[width=5cm]{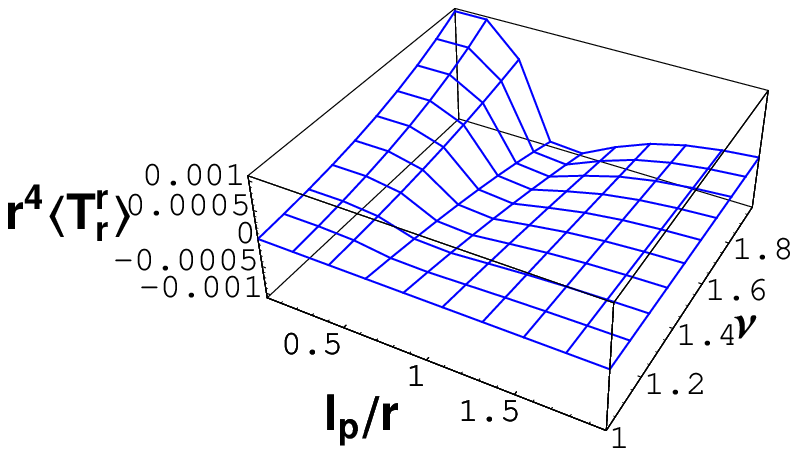}
\\
(P) \hspace{5cm} (A) \hspace{5cm} (S)
\caption{$r^4\langle T_r^r\rangle$ for $\xi=1/6$ is plotted against $l_p/r$ and
$\nu$, for (P) Padmanabhan-type, (A) Abel-type,  and (S) Siegel-type.}
\label{fig5}
\end{figure}

\begin{figure}[ht]
\centering
\includegraphics[width=5cm]{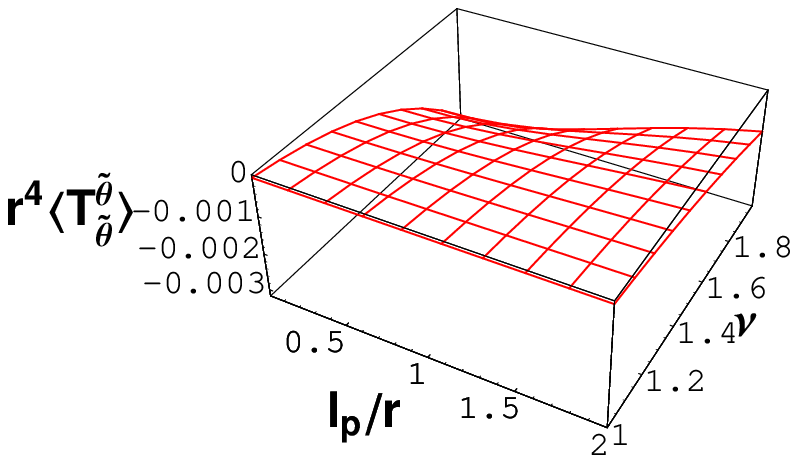}\quad
\includegraphics[width=5cm]{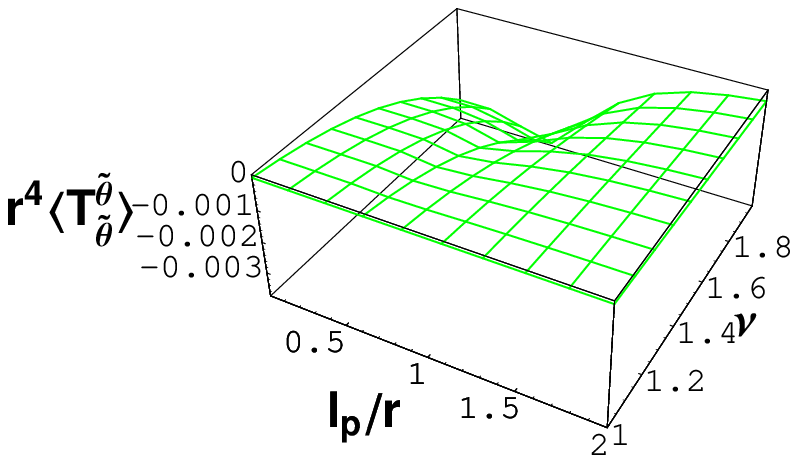}\quad
\includegraphics[width=5cm]{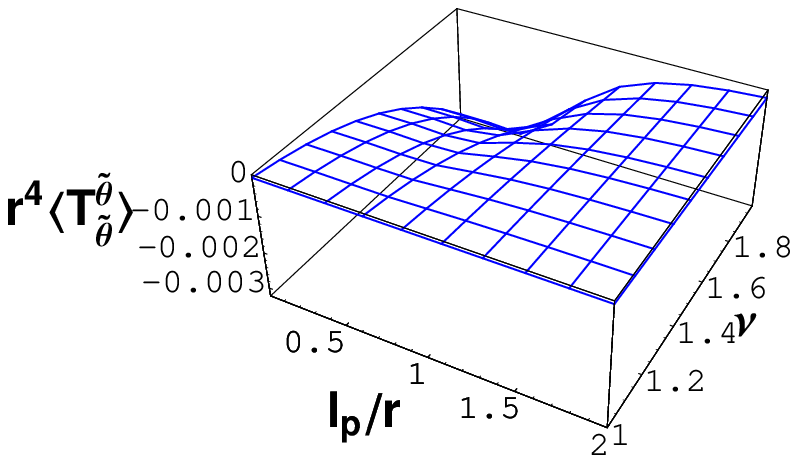}
\\
(P) \hspace{5cm} (A) \hspace{5cm} (S)
\caption{$r^4\langle T_{\tilde{\theta}}^{\tilde{\theta}}\rangle$ for $\xi=1/6$ is
plotted against $l_p/r$ and $\nu$, for (P) Padmanabhan-type, (A) Abel-type,  and
(S) Siegel-type.}
\label{fig6}
\end{figure}

\begin{figure}[ht]
\centering
\includegraphics[width=5cm]{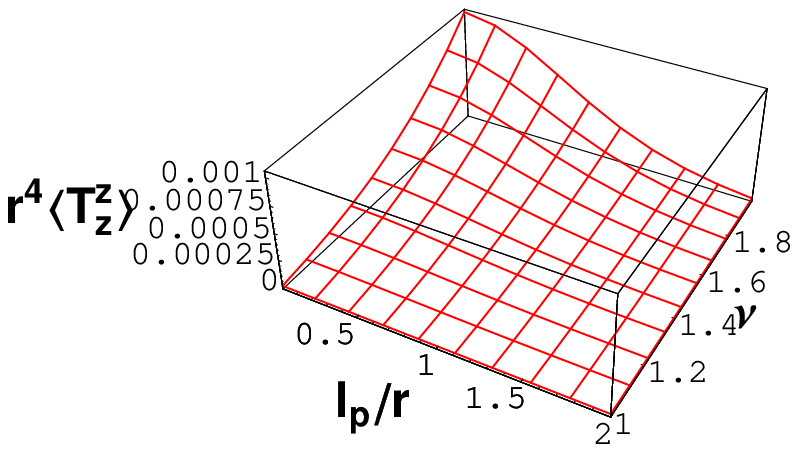}\quad
\includegraphics[width=5cm]{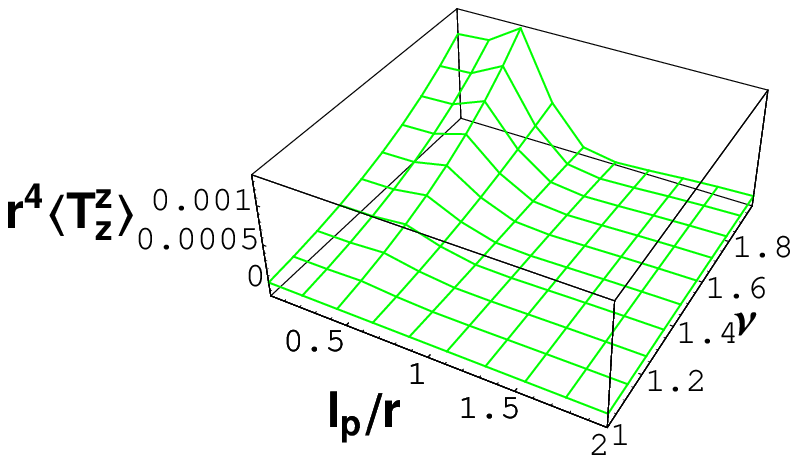}\quad
\includegraphics[width=5cm]{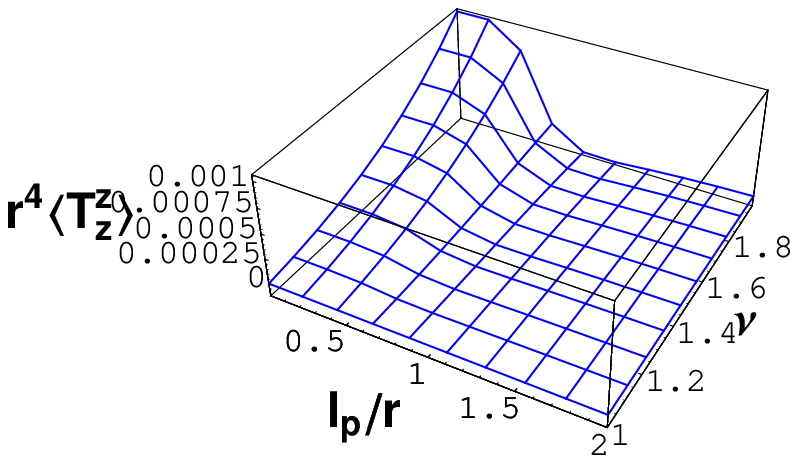}
\\
(P) \hspace{5cm} (A) \hspace{5cm} (S)
\caption{$r^4\langle T_z^z\rangle$ for $\xi=1/6$ is plotted against $l_p/r$ and
$\nu$, for (P) Padmanabhan-type, (A) Abel-type,  and (S) Siegel-type.}
\label{fig7}
\end{figure}

The quantum stress tensors can be obtained from the formula (\ref{qst}) with
(\ref{qstd}) and the Green's functions in the previous subsection.
We exhibit $r^4\langle T_r^r\rangle$ in Fig.~\ref{fig5}, 
$r^4\langle T_{\tilde{\theta}}^{\tilde{\theta}}\rangle$ in Fig.~\ref{fig6}, 
and $r^4\langle T_z^z\rangle$ in Fig.~\ref{fig7}, with a common choice, $\xi=1/6$.
As in the case of $r^2\langle\phi^2\rangle$, quantities of Abel-type
seem to have a small rise in the absolute values around $l_p/r\sim 0.5$.
Except for this feature, Abel-type quantities almost resemble
Siegel-type quantities.
%while the change of values of Padmanabhan's type along with
%$l_p/r$ seems moderate in comparison to other two types.

%%%%%%%%%%%%%%%%%%%%%%%%%%%
% 5
%%%%%%%%%%%%%%%%%%%%%%%%%%%
%\begin{wrapfigure}{r}{5cm}
%\end{wrapfigure}
%%%%%%%%%%%%%%%%%%%%%%%%%%%
%%%%%%%%%%%%%%%%%%%%%%%%%%%
% 6
%%%%%%%%%%%%%%%%%%%%%%%%%%%
%\begin{wrapfigure}{r}{5cm}
%\end{wrapfigure}
%%%%%%%%%%%%%%%%%%%%%%%%%%%
%%%%%%%%%%%%%%%%%%%%%%%%%%%
% 7
%%%%%%%%%%%%%%%%%%%%%%%%%%%
%\begin{wrapfigure}{r}{5cm}
%\end{wrapfigure}
%%%%%%%%%%%%%%%%%%%%%%%%%%%

\begin{figure}[ht]
\centering
\includegraphics[width=5cm]{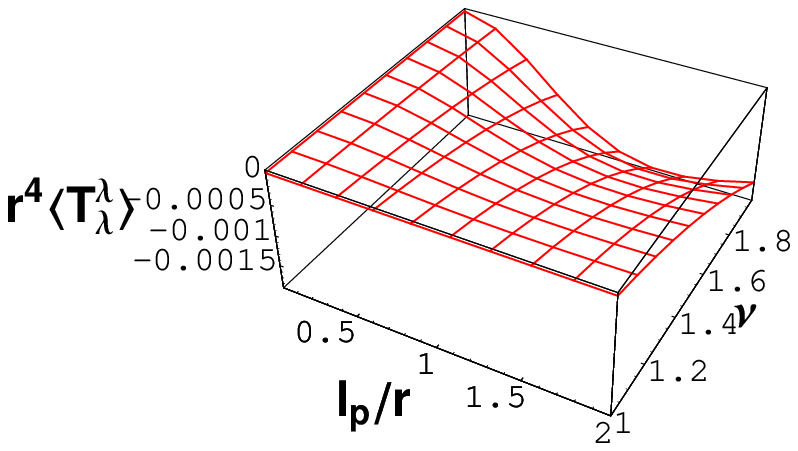}\quad
\includegraphics[width=5cm]{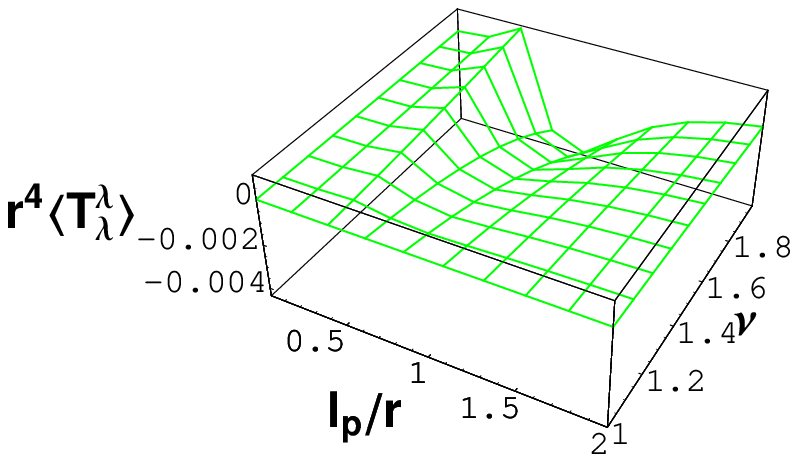}\quad
\includegraphics[width=5cm]{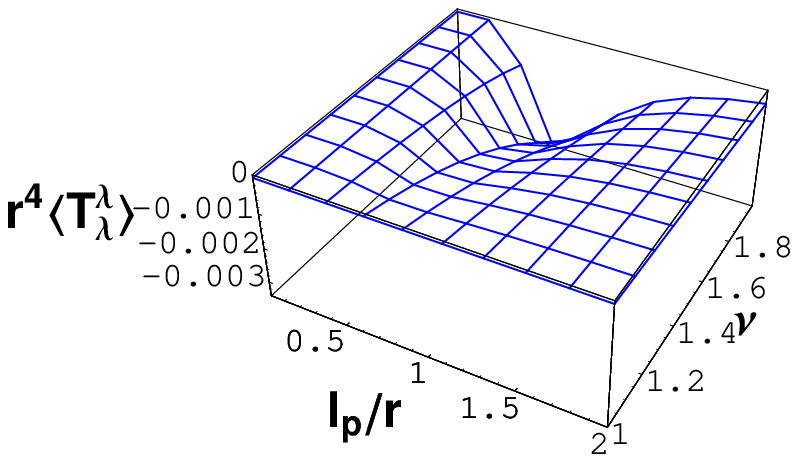}
\\
(P) \hspace{5cm} (A) \hspace{5cm} (S)
\caption{$r^4\langle T_\lambda^\lambda\rangle$ for $\xi=1/6$ is plotted against
$l_p/r$ and
$\nu$, for (P) Padmanabhan-type, (A) Abel-type,  and (S) Siegel-type.}
\label{fig8}
\end{figure}

The trace of the quantum stress tensor does not vanish for finite $l_p$ even if
$\xi=1/6$ in each case. Fig.~\ref{fig8} shows $r^4\langle T_\lambda^\lambda\rangle$
for $\xi=1/6$ in each case. We find, of course, $r^4\langle
T_\lambda^\lambda\rangle=0$ if $l_p=0$ in each case.

%%%%%%%%%%%%%%%%%%%%%%%%%%%
% 8
%%%%%%%%%%%%%%%%%%%%%%%%%%%
%\begin{wrapfigure}{r}{5cm}
%\end{wrapfigure}
%%%%%%%%%%%%%%%%%%%%%%%%%%%

\begin{figure}[ht]
\centering
\includegraphics[width=5cm]{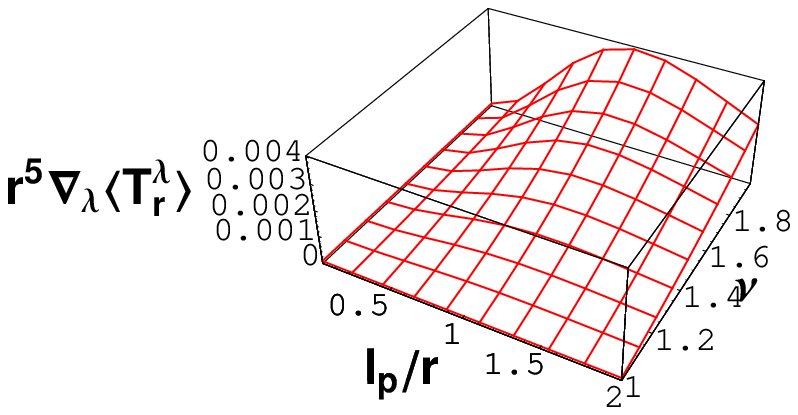}\quad
\includegraphics[width=5cm]{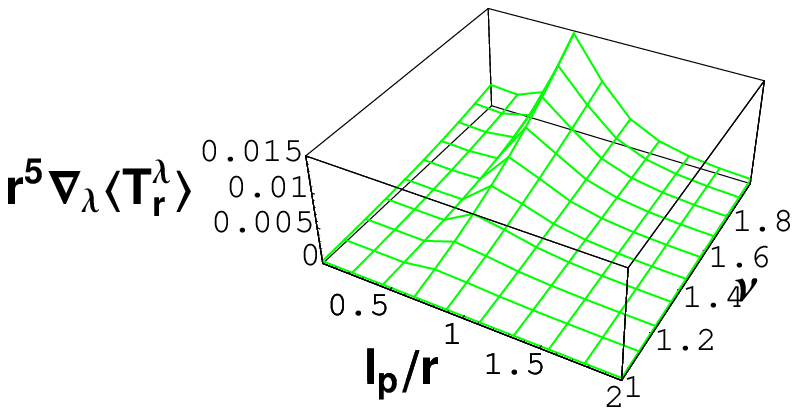}\quad
\includegraphics[width=5cm]{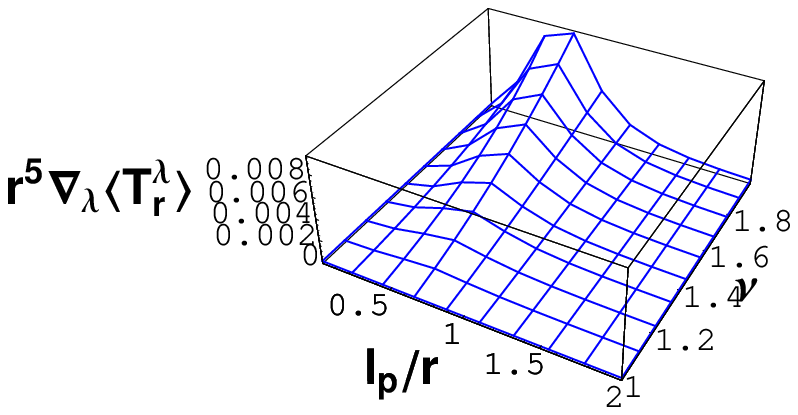}
\\
(P) \hspace{5cm} (A) \hspace{5cm} (S)
\caption{$r^5\nabla_\lambda\langle T_\lambda^\lambda\rangle$ for $\xi=1/6$ is
plotted against
$l_p/r$ and
$\nu$, for (P) Padmanabhan-type, (A) Abel-type,  and (S) Siegel-type.}
\label{fig9}
\end{figure}

Inclusion of $l_p$ tinily violates conservation law $\nabla_\lambda\langle
T_\sigma^\lambda\rangle=0$. Fig.~\ref{fig9} shows $r^5\nabla_\lambda\langle
T_r^\lambda\rangle$ for $\xi=1/6$ in each case. We naturally find that this value
vanishes in each case if
$l_p=0$.

%%%%%%%%%%%%%%%%%%%%%%%%%%%
% 9
%%%%%%%%%%%%%%%%%%%%%%%%%%%
%\begin{wrapfigure}{r}{5cm}
%\end{wrapfigure}
%%%%%%%%%%%%%%%%%%%%%%%%%%%

%%%%%%%%%%%%%%%%%%%%%%%%%%%%%%%%%%%%%%%%%%%%%%%%%%%%%%%%%%%%%%%%%%%%%%%%%%%
%%%%%%%%%%%%%%%%%%%%%%%%%%%%%%%%%%%%%%%%%%%%%%%%%%%%%%%%%%%%%%%%%%%%%%%%%%%
\subsection{The expectation values in the neighborhood of the origin $r=0$}
%%%%%%%%%%%%%%%%%%%%%%%%%%%%%%%%%%%%%%%%%%%%%%%%%%%%%%%%%%%%%%%%%%%%%%%%%%%
%%%%%%%%%%%%%%%%%%%%%%%%%%%%%%%%%%%%%%%%%%%%%%%%%%%%%%%%%%%%%%%%%%%%%%%%%%%

In this subsection, we would like to investigate the values of
$\langle\phi^2\rangle$ and $\langle T_\sigma^\rho\rangle$
in three UV improved schemes
near and at the origin $r=0$, where a conical singularity is located.
Recall that, in the standard scheme without fundamental length,
they behave
$\langle\phi^2\rangle\propto 1/r^2$ and $\langle T_\sigma^\rho\rangle\propto 1/r^4$
in four dimensions, thus the values of them diverge at the origin.

\begin{figure}[ht]
\centering
\includegraphics[width=5cm]{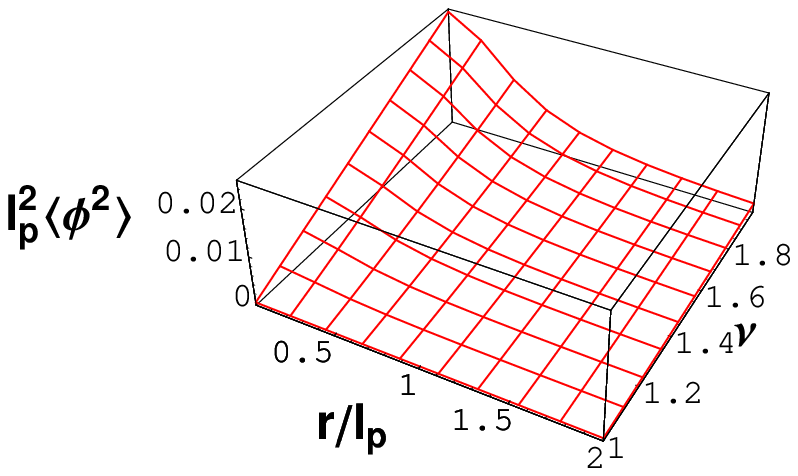}\quad
\includegraphics[width=5cm]{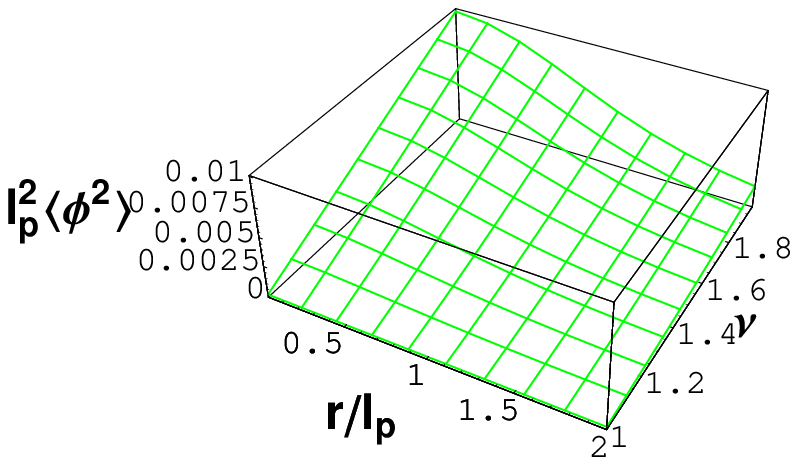}\quad
\includegraphics[width=5cm]{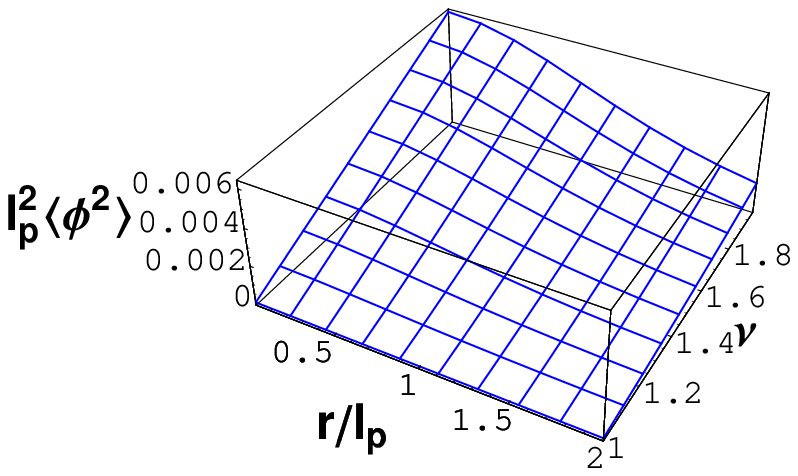}
\\
(P) \hspace{5cm} (A) \hspace{5cm} (S)
\caption{$l_p^2\langle\phi^2\rangle$ is plotted against $r/l_p$ and $\nu$, for (P)
Padmanabhan-type, (A) Abel-type,  and (S) Siegel-type.}
\label{fig10}
\end{figure}

In Fig.~\ref{fig10}, the values of $l_p^2\langle\phi^2\rangle$ are plotted against
$r/l_p$ and $\nu$ for three cases of Green's functions.
We find that $\langle\phi^2\rangle$ is finite at $r=0$ and a monotonic function of
both $r$ and $\nu$ for each type.

%%%%%%%%%%%%%%%%%%%%%%%%%%%
% 10
%%%%%%%%%%%%%%%%%%%%%%%%%%%
%\begin{wrapfigure}{r}{5cm}
%\end{wrapfigure}
%%%%%%%%%%%%%%%%%%%%%%%%%%%

The limiting value $\langle\phi^2\rangle(0)$ can be obtained in a rigorous form
for each type, even in general dimensions. 
To this end, we remark, from (\ref{sk}),
\begin{equation}
\bar{K}_{D,\nu}(0,0;s)={\textstyle
\frac{1}{2\pi(4\pi
s)^{D/2}}}\int_{0}^\infty\left[{\textstyle
-\frac{2\nu\sin\nu\pi}{\cosh\nu
v-\cos\nu\pi}}\right]dv=\frac{\nu-1}{(4\pi
s)^{D/2}}=(\nu-1)K_D(0,0;s)\,,
\end{equation}
and we find
\begin{equation}
\langle\phi^2\rangle^{(*)}(0)=(\nu-1)G_D^{(*)}(0,0)\qquad ({\textstyle(*)=(P),
(A), \mbox{~and~} (S)})\,.
\end{equation}
Here, we should recall that $G_D^{(*)}(x,x')$ is the Green's function in the space
without singularity. According to (\ref{acs}), the values for $D=4$ are found to be
\begin{equation}
G_4^{(P)}(0,0)=\frac{1}{4\pi^2l_p^2}\,,\quad
G_4^{(A)}(0,0)=\frac{1}{32\pi l_p^2}\,,\quad
G_4^{(S)}(0,0)=\frac{1}{16\pi^2l_p^2}\,.
\end{equation}
Therefore, the value of vacuum fluctuation from Padmanabhan-type Green's
function in four dimensions can be confirmed at the special point $r=0$
by using (\ref{vf1}) with (\ref{vf2}).

%%%%%%%%%%%%%%%%%%%%%%%%%%%%%%%%%%%%%%%%%%%%%%%%%%%%%%%%%%%%%%%%%%%%%%%%%%%

\begin{figure}[ht]
\centering
\includegraphics[width=5cm]{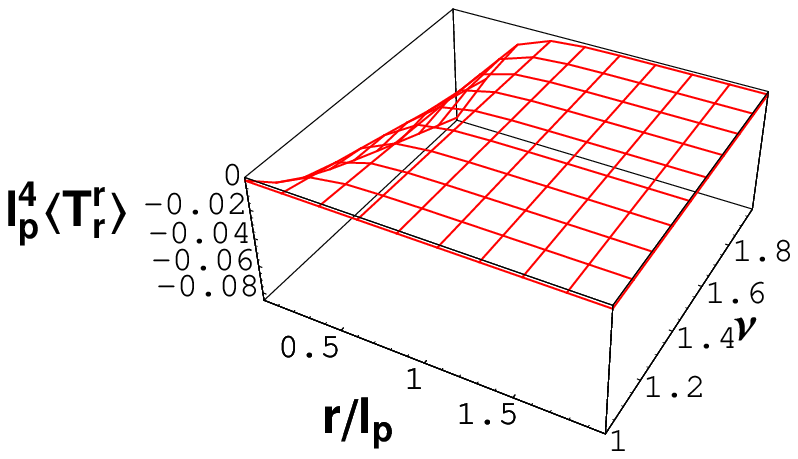}\quad
\includegraphics[width=5cm]{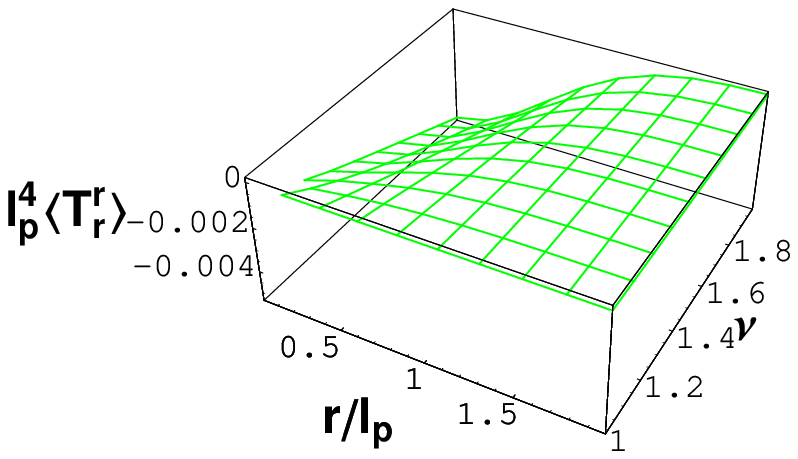}\quad
\includegraphics[width=5cm]{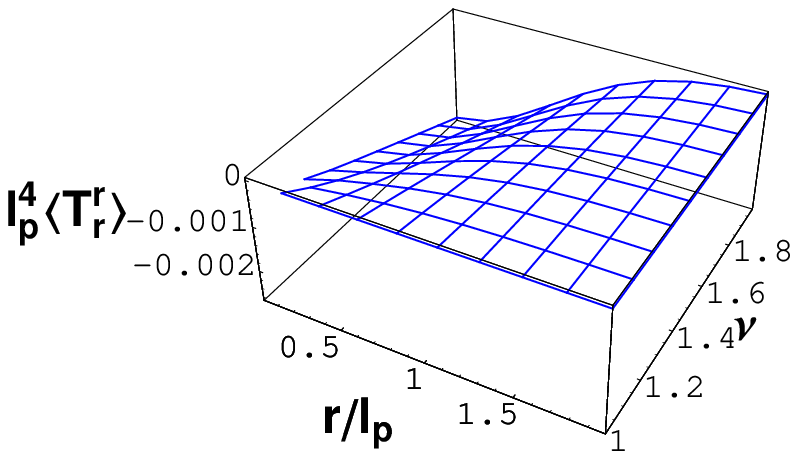}
\\
(P) \hspace{5cm} (A) \hspace{5cm} (S)
\caption{$l_p^4\langle T_r^r\rangle$ for $\xi=1/6$ is plotted against $r/l_p$ and
$\nu$, for (P) Padmanabhan-type, (A) Abel-type,  and (S) Siegel-type.}
\label{fig11}
\end{figure}

\begin{figure}[ht]
\centering
\includegraphics[width=5cm]{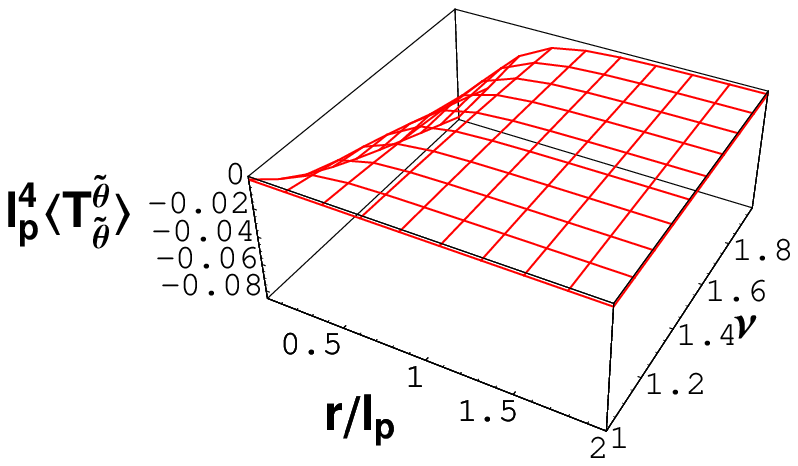}\quad
\includegraphics[width=5cm]{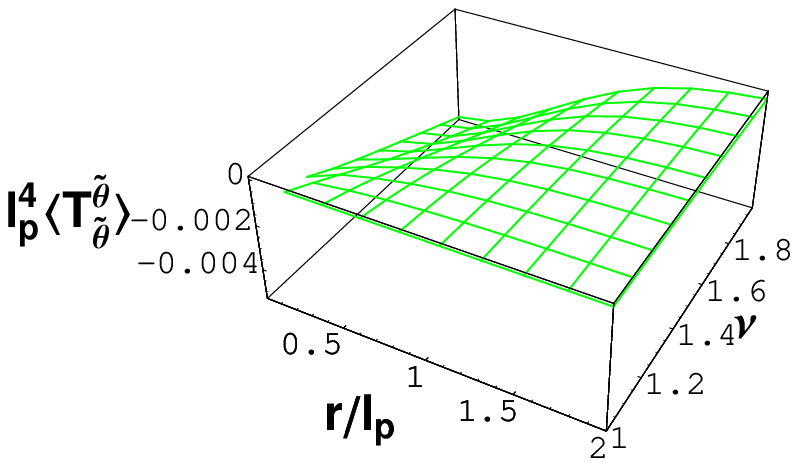}\quad
\includegraphics[width=5cm]{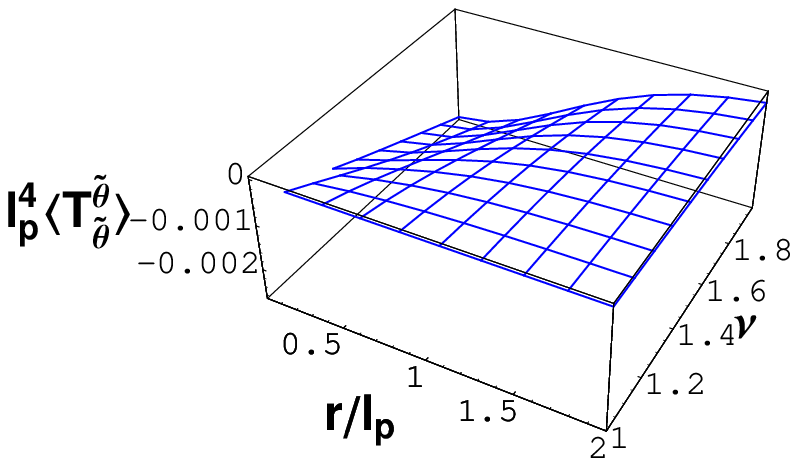}
\\
(P) \hspace{5cm} (A) \hspace{5cm} (S)
\caption{$l_p^4\langle T_{\tilde{\theta}}^{\tilde{\theta}}\rangle$ for $\xi=1/6$ is
plotted against $r/l_p$ and $\nu$, for (P) Padmanabhan-type, (A) Abel-type,  and
(S) Siegel-type.}
\label{fig12}
\end{figure}

\begin{figure}[ht]
\centering
\includegraphics[width=5cm]{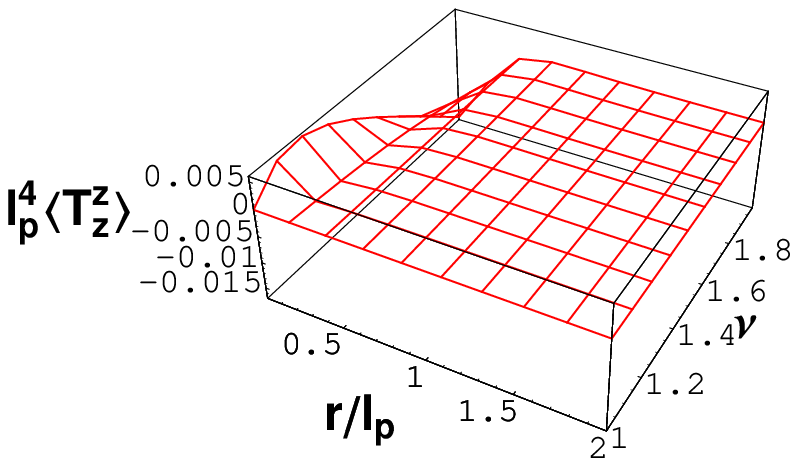}\quad
\includegraphics[width=5cm]{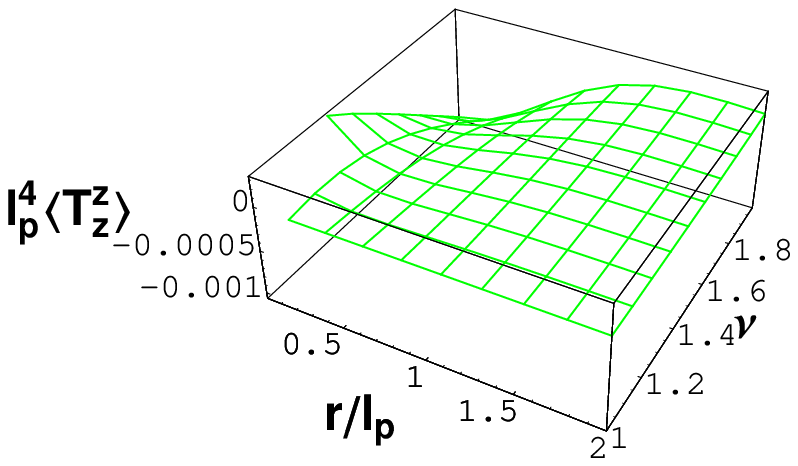}\quad
\includegraphics[width=5cm]{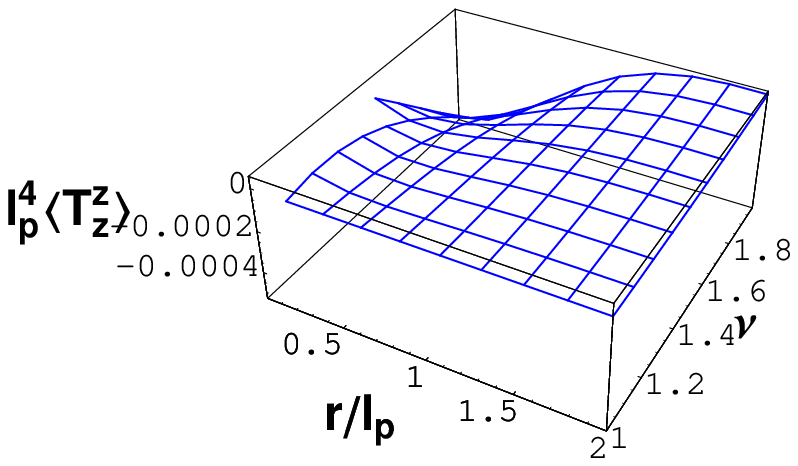}
\\
(P) \hspace{5cm} (A) \hspace{5cm} (S)
\caption{$l_p^4\langle T_z^z\rangle$ for $\xi=1/6$ is plotted against $r/l_p$ and
$\nu$, for (P) Padmanabhan-type, (A) Abel-type,  and (S) Siegel-type.}
\label{fig13}
\end{figure}

\begin{figure}[ht]
\centering
\includegraphics[width=5cm]{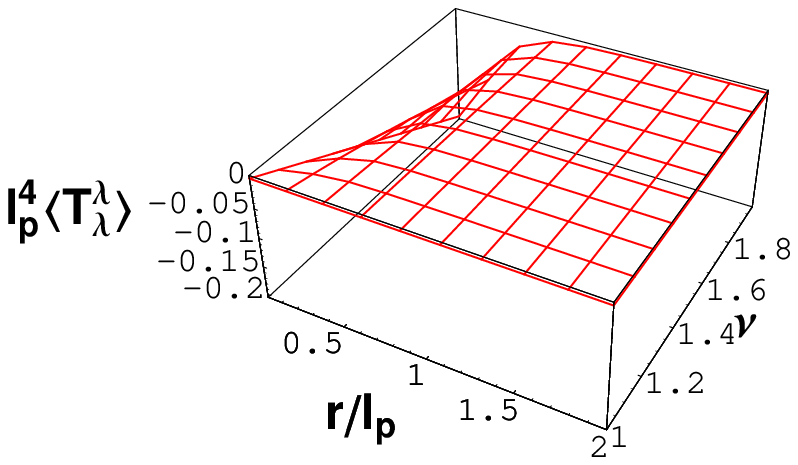}\quad
\includegraphics[width=5cm]{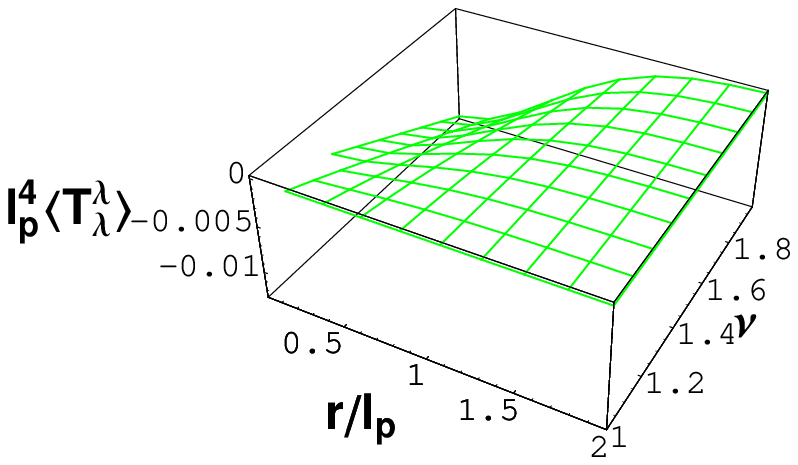}\quad
\includegraphics[width=5cm]{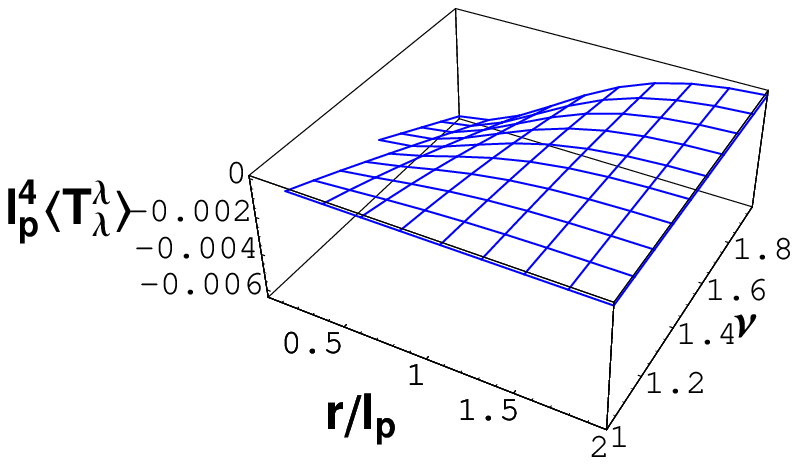}
\\
(P) \hspace{5cm} (A) \hspace{5cm} (S)
\caption{$l_p^4\langle T_\lambda^\lambda\rangle$ for $\xi=1/6$ is plotted against
$r/l_p$ and
$\nu$, for (P) Padmanabhan-type, (A) Abel-type,  and (S) Siegel-type.}
\label{fig14}
\end{figure}

The stress tensors near the origin are shown in Figs.~\ref{fig11}--\ref{fig14}.
We exhibit $l_p^4\langle T_r^r\rangle$ in Fig.~\ref{fig11}, 
$l_p^4\langle T_{\tilde{\theta}}^{\tilde{\theta}}\rangle$ in Fig.~\ref{fig12}, 
$l_p^4\langle T_z^z\rangle$ in Fig.~\ref{fig13}, and $l_p^4\langle
T_\lambda^\lambda\rangle$ in Fig.~\ref{fig14}, with a common choice, $\xi=1/6$.

%%%%%%%%%%%%%%%%%%%%%%%%%%%
% 11
%%%%%%%%%%%%%%%%%%%%%%%%%%%
%\begin{wrapfigure}{r}{5cm}
%\end{wrapfigure}
%%%%%%%%%%%%%%%%%%%%%%%%%%%
%%%%%%%%%%%%%%%%%%%%%%%%%%%
% 12
%%%%%%%%%%%%%%%%%%%%%%%%%%%
%\begin{wrapfigure}{r}{5cm}
%\end{wrapfigure}
%%%%%%%%%%%%%%%%%%%%%%%%%%%
%%%%%%%%%%%%%%%%%%%%%%%%%%%
% 13
%%%%%%%%%%%%%%%%%%%%%%%%%%%
%\begin{wrapfigure}{r}{5cm}
%\end{wrapfigure}
%%%%%%%%%%%%%%%%%%%%%%%%%%%
%%%%%%%%%%%%%%%%%%%%%%%%%%%
% 14
%%%%%%%%%%%%%%%%%%%%%%%%%%%
%\begin{wrapfigure}{r}{5cm}
%\end{wrapfigure}
%%%%%%%%%%%%%%%%%%%%%%%%%%%

The values of $\langle T^\rho_\sigma\rangle^{(*)}(0)$ are also analytically
expressed, similarly to the case with the values of $\langle\phi^2\rangle(0)$,
containing
$G_{D+2}^{(*)}(0,0)$ in the present case, however.%
\footnote{The essential formulas for integration are collected in Appendix
\ref{AB}.}
 That is:
\begin{eqnarray}
\langle T^r_r\rangle^{(*)}(0)&=&\langle
T^{\tilde{\theta}}_{\tilde{\theta}}\rangle^{(*)}(0)=-
4\pi G_{D+2}^{(*)}(0,0)\left[
{\textstyle\frac{(D-2)[D(\nu-1)+1]}{4(D-1)}
+\nu\left(\xi-\frac{D-2}{4(D-1)}\right)}
\right]\,,\\
\langle T^z_z\rangle^{(*)}(0)&=&-
4\pi G_{D+2}^{(*)}(0,0)\left[{\textstyle\frac{D(D-3)(\nu-1)-2}{4(D-1)}
+2\nu\left(\xi-\frac{D-2}{4(D-1)}\right)}\right]\,,\\
\langle T^\lambda_\lambda\rangle^{(*)}(0)&=&-
4\pi G_{D+2}^{(*)}(0,0)\left[{\textstyle\frac{D(D-2)(\nu-1)}{4(D-1)}
+2(D-1)\nu\left(\xi-\frac{D-2}{4(D-1)}\right)}\right]\,,
\end{eqnarray}
where the choice $\xi=\frac{D-2}{4(D-1)}$ is called the conformal coupling in $D$
dimensions.
Incidentally, the values of the coincidence limit of $G_{D+2}^{(*)}(0,0)$ for $D=4$
are found to be
\begin{equation}
G_6^{(P)}(0,0)=\frac{1}{4\pi^3l_p^4}\,,\quad
G_6^{(A)}(0,0)=\frac{1}{64\pi^3l_p^4}\,,\quad
G_6^{(S)}(0,0)=\frac{1}{128\pi^3l_p^4}\,.
\end{equation}
It is interesting to point out that there is the relevance to the Green's function
in ``other dimensions'' and it reminds us of discussion in
Ref.~\cite{CFI}.

We find that quantum stress tensors $\langle
T^\rho_\sigma\rangle$ are finite even if $\nu=1$.%
\footnote{For Padmanabhan's type, it is caused from the singularities in functions
$f_2$ and $f_3$ at $(\nu, r)=(1,0)$.} Although the origin of this ``anomaly''
has not been elucidated yet, this will be discussed later in Sec.~\ref{conclusion}.

%%%%%%%%%%%%%%%%%%%%%%%%%%%%%%%%%%%%%%%%%%%%%%%%%%%%%%%%%%%%%%%%%%%%%%%%%%%

\begin{figure}[ht]
\centering
\includegraphics[width=5cm]{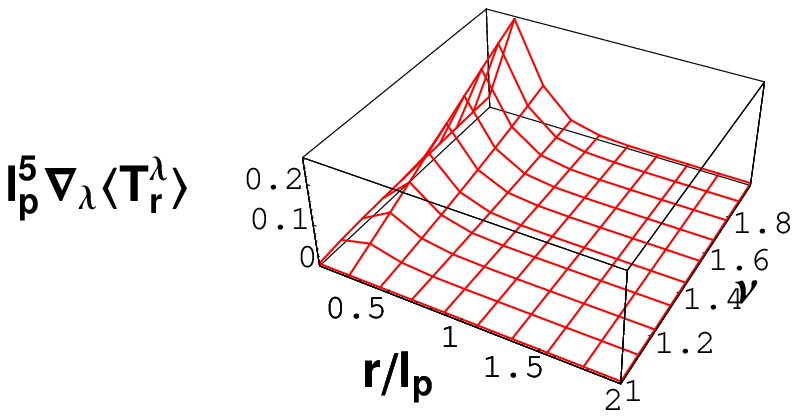}\quad
\includegraphics[width=5cm]{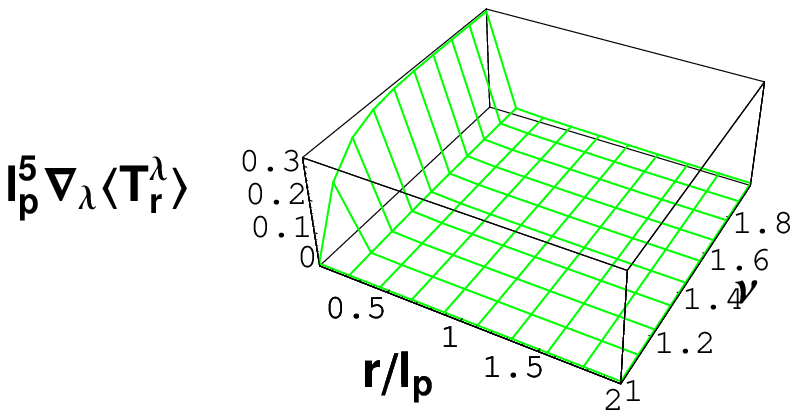}\quad
\includegraphics[width=5cm]{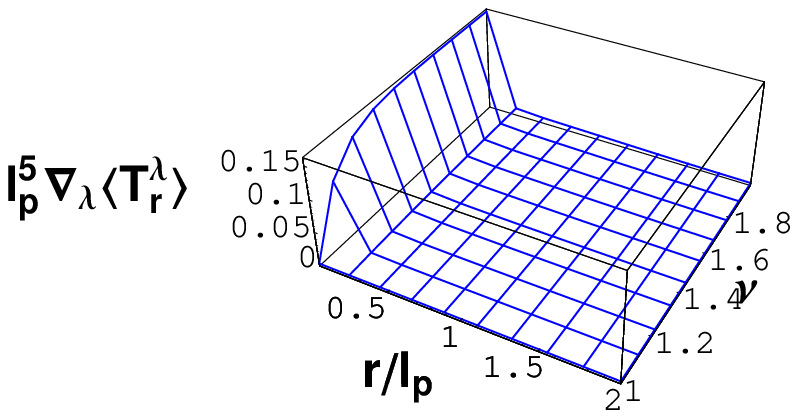}
\\
(P) \hspace{5cm} (A) \hspace{5cm} (S)
\caption{$l_p^5\nabla_\lambda\langle T_r^\lambda\rangle$ for $\xi=1/6$ is
plotted against
$r/l_p$ and
$\nu$, for (P) Padmanabhan-type, (A) Abel-type,  and (S) Siegel-type.}
\label{fig15}
\end{figure}

Finally, we show $l_p^5\nabla_\lambda\langle T_r^\lambda\rangle$ in
Fig.~\ref{fig15} in each case. This value is non-zero near the conical singularity
but rapidly falls off to zero at $r\gg l_p$ in each case.

%%%%%%%%%%%%%%%%%%%%%%%%%%%
% 15
%%%%%%%%%%%%%%%%%%%%%%%%%%%
%\begin{wrapfigure}{r}{5cm}
%\end{wrapfigure}
%%%%%%%%%%%%%%%%%%%%%%

%%%%%%%%%%%%%%%%%%%%%%%%%%%%%%%%%%%%%%%%%%%%%%%%%%%%%%%%%%%%%%%%%%%%%%%%%%%

%%%%%%%%%%%%%%%%%%%%%%%%%%%%%%%%%%%%%%%%%%%%%%%%%%%%%%%%%%%%%%%%%%%%%%%%%%%
%%%%%%%%%%%%%%%%%%%%%%%%%%%%%%%%%%%%%%%%%%%%%%%%%%%%%%%%%%%%%%%%%%%%%%%%%%%
%%%%%%%%%%%%%%%%%%%%%%%%%%%%%%%%%%%%%%%%%%%%%%%%%%%%%%%%%%%%%%%%%%%%%%%%%%%
\section{Conclusion}
\label{conclusion}
%%%%%%%%%%%%%%%%%%%%%%%%%%%%%%%%%%%%%%%%%%%%%%%%%%%%%%%%%%%%%%%%%%%%%%%%%%%
%%%%%%%%%%%%%%%%%%%%%%%%%%%%%%%%%%%%%%%%%%%%%%%%%%%%%%%%%%%%%%%%%%%%%%%%%%%
%%%%%%%%%%%%%%%%%%%%%%%%%%%%%%%%%%%%%%%%%%%%%%%%%%%%%%%%%%%%%%%%%%%%%%%%%%%

We have presented the vacuum expectation values for a massless scalar field
obtained from three types of the UV improved Green's functions
in non-simply connected spaces. Although the behavior of these values in a various
range of scale is almost common, the quantities calculated from Abel-type Green's
function show a minute behavior if the typical scale of the system is close to
the Planck scale, i.e., the cutoff scale.
Interestingly, the present results should be directly relevant to the study of
the Unruh effect \cite{Unruh,Takagi,Dowker} and the quantum inconsistency of the
space with closed timelike curves \cite{Hawking}.

The UV improved Green's functions we examined in this paper are mathematically
simple ones, where the Planck length is introduced ``by hand'', though certain
physical motivation exists for each type of UV modification. 
Through the present study, however, it is revealed that the quantum quantities have
slightly different behaviors around the small scale according to the type of
UV completion. These results would be useful when we pursue the fundamental
origin of UV completion and when we consider the back-reaction of quantum
effects to space(time) structure.
In any case, the mechanism of UV completion will require further investigation.

The ``anomaly'' encountered in the quantum stress tensors just at the conical
singularity  should be investigated further. This should be linked to the
prescription of ``refinement'' or ``renormalization'' of the quantum quantities.
The finite but comparatively large values of order $l_p^{-n}$ ($n$ is a positive
integer) in quantum corrections have not ever been observed and should not directly
affect known physical consequence. Anyway, the ``renormalization''
such as $\overline{\langle T_{\rho\sigma}\rangle}=\langle
T_{\rho\sigma}\rangle-\lim_{\nu\rightarrow 1}\langle
T_{\rho\sigma}\rangle$ can also be allowed, so we think that it would be
necessary to study the connection between the renormalization and the
back-reaction problem, where we should reconsider coupling to space(time)
geometry. 

Generalizing to other situations should also be possible.
We have restricted our attention to the massless scalar case in this paper.
The extension of the present analyses to massive cases should be straightforward,
at least for the models of Padmanabhan-type and Siegel-type.
It would be interesting to generalize our study to calculations in curved spaces
or near black holes, including scattering problems with UV completion.
Although the general treatment of the curved background might be hard,
we may start with adopting a perturvative expansion around a trivial space.

Alternatively, the most direct generalization is considering constant-curvature
background spaces, such as $S^N$ corresponding to Euclidean de Sitter space and  
$H^N$ corresponding to Euclidean anti-de Sitter space. The standard heat kernel
in such spaces are already known (see, for example, Ref.~\cite{Camporesi}).
Accordingly, the quantum effects around BTZ black holes
\cite{Steif,LO,SM1,SM2}
with UV completion are within the scope of feasible study in near future.

We consider that mathematical properties of Green's functions and heat kernels with
the cutoff scale is also an interesting subject to study.
We have already seen that the conservation law and the conformal symmetry are
violated and broken by introducing the Planck length in three cases studied in this
paper. In an academic point of view, we should study where, when, and how such
fundamental law and symmetry can be protected in more general way of UV completion
with close inspection. That is to say, we should take corrections in the
definition of the stress tensor and field equations into account. In addition, we
intend to investigate the mathematical nature of the heat kernel in the UV
completion schemes. For example, we notice the fact that the standard heat kernel
without cutoff scale in a direct-product space is the product of the heat kernels
associated to two spaces. The fundamental ``rule'' in this level may yield a new
guideline in theoretical research of physical contribution from very small scale
physics.

%%%%%%%%%%%%%%%%%%%%%%%%%%%%%%%%%%%%%%%%%%%%%%%%%%%%%%%%%%%%%%%%%
%%%%%%%%%%%%%%%%%%%%%%%%%%%%%%%%%%%%%%%%%%%%%%%%%%%%%%%%%%%%%%%%%
\appendix
%%%%%%%%%%%%%%%%%%%%%%%%%%%%%%%%%%%%%%%%%%%%%%%%%%%%%%%%%%%%%%%%%

%%%%%%%%%%%%%%%%%%%%%%%%%%%%%%%%%%%%%%%%%%%%%%%%%%%%%%%%%%%%%%%%%%%%%%%%%%%
%%%%%%%%%%%%%%%%%%%%%%%%%%%%%%%%%%%%%%%%%%%%%%%%%%%%%%%%%%%%%%%%%%%%%%%%%%%
%%%%%%%%%%%%%%%%%%%%%%%%%%%%%%%%%%%%%%%%%%%%%%%%%%%%%%%%%%%%%%%%%%%%%%%%%%%
\section{definitions of special functions and their properties}\label{AA}
%%%%%%%%%%%%%%%%%%%%%%%%%%%%%%%%%%%%%%%%%%%%%%%%%%%%%%%%%%%%%%%%%%%%%%%%%%%
%%%%%%%%%%%%%%%%%%%%%%%%%%%%%%%%%%%%%%%%%%%%%%%%%%%%%%%%%%%%%%%%%%%%%%%%%%%
%%%%%%%%%%%%%%%%%%%%%%%%%%%%%%%%%%%%%%%%%%%%%%%%%%%%%%%%%%%%%%%%%%%%%%%%%%%
Almost all definitions and properties of the special functions exhibited below can
be found in Ref.~\cite{GR}.
%%%%%%%%%%%%%%%%%%%%%%%%%%%%%%%%%%%%%%%%%%%%%%%%%%%%%%%%%%%%%%%%%%%%%%%%%%%
%%%%%%%%%%%%%%%%%%%%%%%%%%%%%%%%%%%%%%%%%%%%%%%%%%%%%%%%%%%%%%%%%%%%%%%%%%%
\subsection{Modified Bessel function of the first kind}
%%%%%%%%%%%%%%%%%%%%%%%%%%%%%%%%%%%%%%%%%%%%%%%%%%%%%%%%%%%%%%%%%%%%%%%%%%%
%%%%%%%%%%%%%%%%%%%%%%%%%%%%%%%%%%%%%%%%%%%%%%%%%%%%%%%%%%%%%%%%%%%%%%%%%%%
\begin{eqnarray}
I_\nu(z)&=&\frac{2(z/2)^\nu}{\sqrt{\pi}\,\Gamma({\textstyle\nu+\frac{1}{2}})}
\int_0^{\pi/2}\cosh(z\cos\theta)\sin^{2\nu}\theta d\theta\nonumber \\
&=&\frac{1}{2\pi}\int_{-\pi}^\pi e^{z\cos\theta}\cos\nu\theta\,d\theta
-\frac{\sin\nu\pi}{\pi}\int_0^\infty
e^{-z\cosh v-\nu v}\,dv\,.
\end{eqnarray}

The integral formula
\begin{equation}
\int_0^\infty e^{-px}I_\nu(cx)dx=\frac{c^\nu}{\sqrt{p^2-c^2}(p+\sqrt{p^2-c^2})^\nu}
\end{equation}
leads to (\ref{cf}) from (\ref{conK}).
%%%%%%%%%%%%%%%%%%%%%%%%%%%%%%%%%%%%%%%%%%%%%%%%%%%%%%%%%%%%%%%%%%%%%%%%%%%
%%%%%%%%%%%%%%%%%%%%%%%%%%%%%%%%%%%%%%%%%%%%%%%%%%%%%%%%%%%%%%%%%%%%%%%%%%%
\subsection{Modified Struve function}
%%%%%%%%%%%%%%%%%%%%%%%%%%%%%%%%%%%%%%%%%%%%%%%%%%%%%%%%%%%%%%%%%%%%%%%%%%%
%%%%%%%%%%%%%%%%%%%%%%%%%%%%%%%%%%%%%%%%%%%%%%%%%%%%%%%%%%%%%%%%%%%%%%%%%%%
\begin{equation}
\mathbf{L}_\nu(z)=\frac{2(z/2)^\nu}{\sqrt{\pi}\,\Gamma({\textstyle\nu+\frac{1}{2}})}
\int_0^{\pi/2}\sinh(z\cos\theta)\sin^{2\nu}\theta d\theta\,.
\end{equation}

%%%%%%%%%%%%%%%%%%%%%%%%%%%%%%%%%%%%%%%%%%%%%%%%%%%%%%%%%%%%%%%%%%%%%%%%%%%
%%%%%%%%%%%%%%%%%%%%%%%%%%%%%%%%%%%%%%%%%%%%%%%%%%%%%%%%%%%%%%%%%%%%%%%%%%%
\subsection{Jacobi's theta function}
%%%%%%%%%%%%%%%%%%%%%%%%%%%%%%%%%%%%%%%%%%%%%%%%%%%%%%%%%%%%%%%%%%%%%%%%%%%
%%%%%%%%%%%%%%%%%%%%%%%%%%%%%%%%%%%%%%%%%%%%%%%%%%%%%%%%%%%%%%%%%%%%%%%%%%%
\begin{equation}
\vartheta_3(v,\tau)=1+2\sum_{n=1}^\infty e^{\tau\pi i n^2}\cos(2n\pi v)
=e^{\pi i/4}\tau^{-1/2}e^{-\pi i v^2/\tau}\vartheta_3(v/\tau,-1/\tau)\,.
\end{equation}
Consequently, one can find
\begin{equation}
\sum_{n=-\infty}^\infty e^{-4\pi^2(n+\delta)^2s/L^2}e^{2\pi i(n+\delta)y/L}=
\frac{L}{\sqrt{4\pi s}}\sum_{n=-\infty}^\infty e^{-(y-nL)^2/(4s)+2\pi\delta i n}\,.
\end{equation}

%%%%%%%%%%%%%%%%%%%%%%%%%%%%%%%%%%%%%%%%%%%%%%%%%%%%%%%%%%%%%%%%%%%%%%%%%%%
%%%%%%%%%%%%%%%%%%%%%%%%%%%%%%%%%%%%%%%%%%%%%%%%%%%%%%%%%%%%%%%%%%%%%%%%%%%
\subsection{Modified Bessel function of the second kind}
%%%%%%%%%%%%%%%%%%%%%%%%%%%%%%%%%%%%%%%%%%%%%%%%%%%%%%%%%%%%%%%%%%%%%%%%%%%
%%%%%%%%%%%%%%%%%%%%%%%%%%%%%%%%%%%%%%%%%%%%%%%%%%%%%%%%%%%%%%%%%%%%%%%%%%%
\begin{eqnarray}
K_{\nu}(z)&=&K_{-\nu}(z)=\frac{1}{2}\left(\frac{z}{2}\right)^\nu\int_0^\infty
\exp\left[-t-\frac{z^2}{4t}\right]t^{-\nu-1} dt\nonumber \\
&=&\frac{1}{2}\int_{-\infty}^\infty
\exp\left[-\nu t-z\cosh t\right] dt
=\int_{0}^\infty
\exp\left[-z\cosh t\right]\cosh\nu t\, dt\,.
\end{eqnarray}

\begin{equation}
K_{\nu}(z)\sim\frac{2^{\nu-1}\Gamma(\nu)}{z^\nu}\quad z\rightarrow 0\,,
\qquad
K_{\nu}(z)\sim\sqrt{\frac{\pi}{2z}}e^{-z}\quad z\rightarrow \infty\,.
\end{equation}

%%%%%%%%%%%%%%%%%%%%%%%%%%%%%%%%%%%%%%%%%%%%%%%%%%%%%%%%%%%%%%%%%%%%%%%%%%%
%%%%%%%%%%%%%%%%%%%%%%%%%%%%%%%%%%%%%%%%%%%%%%%%%%%%%%%%%%%%%%%%%%%%%%%%%%%
\subsection{Summation formulas}
%%%%%%%%%%%%%%%%%%%%%%%%%%%%%%%%%%%%%%%%%%%%%%%%%%%%%%%%%%%%%%%%%%%%%%%%%%%
%%%%%%%%%%%%%%%%%%%%%%%%%%%%%%%%%%%%%%%%%%%%%%%%%%%%%%%%%%%%%%%%%%%%%%%%%%%
\begin{equation}
\sum_{n=1}^\infty\frac{\cos(n\delta)}{n^2+a^2}=\frac{\pi}{2a}
\frac{\cosh[a(\pi-\delta)]}{\sinh(a\pi)}-\frac{1}{2a^2}\,.
\end{equation}
%%%%%%%%%%%%%%%%%%%%%%%%%%%%%%%%%%%%%%%%%%%%%%%%%%%%%%%%%%%%%%%%%%%%%%%%%%%
\begin{equation}
\sum_{n=1}^\infty\frac{\cos(n\delta)}{(n^2+a^2)^2}=-\frac{1}{2a^4}+
\frac{\pi}{4a^3}
\frac{\cosh[a(\pi-\delta)]+a\delta\sinh[a(\pi-\delta)]}{\sinh(a\pi)}+\frac{\pi^2}{4a^2}
\frac{\cosh(a\delta)}{\sinh^2(a\pi)}\,.
\end{equation}
%%%%%%%%%%%%%%%%%%%%%%%%%%%%%%%%%%%%%%%%%%%%%%%%%%%%%%%%%%%%%%%%%%%%%%%%%%%

%%%%%%%%%%%%%%%%%%%%%%%%%%%%%%%%%%%%%%%%%%%%%%%%%%%%%%%%%%%%%%%%%%%%%%%%%%%
%%%%%%%%%%%%%%%%%%%%%%%%%%%%%%%%%%%%%%%%%%%%%%%%%%%%%%%%%%%%%%%%%%%%%%%%%%%
%%%%%%%%%%%%%%%%%%%%%%%%%%%%%%%%%%%%%%%%%%%%%%%%%%%%%%%%%%%%%%%%%%%%%%%%%%%
\section{integration formulas}\label{AB}
%%%%%%%%%%%%%%%%%%%%%%%%%%%%%%%%%%%%%%%%%%%%%%%%%%%%%%%%%%%%%%%%%%%%%%%%%%%
%%%%%%%%%%%%%%%%%%%%%%%%%%%%%%%%%%%%%%%%%%%%%%%%%%%%%%%%%%%%%%%%%%%%%%%%%%%
%%%%%%%%%%%%%%%%%%%%%%%%%%%%%%%%%%%%%%%%%%%%%%%%%%%%%%%%%%%%%%%%%%%%%%%%%%%
%%%%%%%%%%%%%%%%%%%%%%%%%%%%%%%%%%%%%%%%%%%%%%%%%%%%%%%%%%%%%%%%%%%%%%%%%%%
The formulas below have been used in calculations of quantum expectation values in
the limit $r\rightarrow 0$ in Sec.~\ref{sec4}.
\begin{equation}
\int_0^\infty\frac{-2\nu\sin\nu\pi}{\cosh\nu v-\cos\nu\pi}dv=2\pi(\nu-1)
\quad(\nu>1)\,,
\end{equation}
\begin{equation}
\int_0^\infty\frac{-2\nu\sin\nu\pi}{\cosh\nu v-\cos\nu\pi}\cosh v\,dv=2\pi
\quad(\nu>1)\,,
\end{equation}
\begin{equation}
\int_0^\infty\frac{\nu^3\sin\nu\pi(-3+2\cos\nu\pi\cosh\nu v+\cosh 2\nu
v)}{(\cosh\nu v-\cos\nu\pi)^3}dv=0
\quad(\nu>1)\,,
\end{equation}
\begin{equation}
\int_0^\infty\frac{\nu^3\sin\nu\pi(-3+2\cos\nu\pi\cosh\nu v+\cosh 2\nu
v)}{(\cosh\nu v-\cos\nu\pi)^3}\cosh
v\, dv=2\pi
\quad(\nu>1)\,.
\end{equation}

%%%%%%%%%%%%%%%%%%%%%%%%%%%%%%%%%%%%%%%%%%%%%%%%%%%%%%%%%%%%%%%%%%%%%%%%%%%
\acknowledgments
%%%%%%%%%%%%%%%%%%%%%%%%%%%%%%%%%%%%%%%%%%%%%%%%%%%%%%%%%%%%%%%%%%%%%%%%%%%
%Acknowledgements
%%%%%%%%%%%%%%%%%%%%%%%%%%%%%%%%%%%%%%%%%%%%%%%%%%%%%%%%%%%%%%%%%%%%%%%%%%%
%\begin{acknowledgments}
The authors are grateful to Ryo Saito for useful discussions.
%the organizers of JGRG21, where our
%partial result %({\tt [arXiv:10mm.xxxx]}) 
%was presented. %for elucidating comments.
%This study is supported in part by the Grant-in-Aid of Nikaido Research 
%Fund.
%\end{acknowledgments}
%%%%%%%%%%%%%%%%%%%%%%%%%%%%%%%%%%%%%%%%%%%%%%%%%%%%%%%%%%%%%%%%%%%%%%%%%%%

%%%%%%%%%%%%%%%%%%%%%%%%%%%%%%%%%%%%%%%%%
%%%%%%%%%%%%%%%%%%%%%%%%%%%%%%%%%%%%%%%%%
%%%
%%%   References
%%%
%%%%%%%%%%%%%%%%%%%%%%%%%%%%%%%%%%%%%%%%%
%%%%%%%%%%%%%%%%%%%%%%%%%%%%%%%%%%%%%%%%%
%%%%%%%%%%%%%%%%%%%%%%%%%%%%%%%%%%%%%%%%%%%%%%%%%%%%%%%%%%%%%%%%%%%%%%%%%%%
%thebibliography
%%%%%%%%%%%%%%%%%%%%%%%%%%%%%%%%%%%%%%%%%%%%%%%%%%%%%%%%%%%%%%%%%%%%%%%%%%%
%\bibliographystyle{apsrev}
\bibliographystyle{apsrev4-1}
%\bibliography{}

%%%%%%%%%%%%%%%%%%%%%%%%%%%%%%%%%%%%%%%%%%%%%%%%%%%%%%%%%%%%%%%%%%%%%%%%%%%
%%%%%%%%%%%%%%%%%%%%%%%%%%%%%%%%%%%%%%%%%%%%%%%%%%%%%%%%%%%%%%%%%%%%%%%%%%%
%%%%%%%%%%%%%%%%%%%%%%%%%%%%%%%%%%%%%%%%%%%%%%%%%%%%%%%%%%%%%%%%%%%%%%%%%%
\end{document}